\documentclass[letterpaper, 10pt, conference]{IEEEtran}  

\IEEEoverridecommandlockouts                              

\usepackage{graphicx} 
\usepackage{epstopdf} 
\usepackage{amsmath} 
\usepackage{amssymb}  
\usepackage{subcaption}

\usepackage{amsthm,amsfonts}
\usepackage{xcolor}
\usepackage{algorithm}
\usepackage{algorithmic}
\usepackage{tcolorbox}
\usepackage[english]{babel}
\usepackage{blindtext}

\newtheorem{theorem}{Theorem}

\theoremstyle{remark}
\newtheorem{remark}{Remark}
\newtheorem{problem}{Problem}
\newtheorem{definition}{Definition}
\newtheorem{assume}{Assumption}
\newtheorem{property}{Property}

\title{Three Dimensional Moving Path Following Control for Robotic Vehicles with Minimum Positive Forward Speed}

\author{R. Praveen Jain, Jo\~{a}o Borges de Sousa, and A. Pedro Aguiar 
\thanks{This project was supported by the project IMPROVE - POCI- 01-0145- FEDER-031823 - funded by FEDER funds through COMPETE2020 - POCI and by national funds (PIDDAC).}
\thanks{R. Praveen Jain is with Department of Engineering Cybernetics, Norwegian University of Science and Technology, 7491 Trondheim, Norway, and A. Pedro Aguiar and Jo\~{a}o Borges de Sousa are with the Department of Electrical and Computer Engineering, Faculty of Engineering, University of Porto, 4200-465 Porto, Portugal.
        {\tt \small ravinder.p.k.jain@ntnu.no, \{pedro.aguiar, jtasso\}@fe.up.pt}}%
}

\begin{document}

\maketitle

\begin{abstract}
This paper addresses the problem of steering a robotic vehicle along a geometric path specified with respect to a reference frame moving in three dimensions, termed the Moving Path Following (MPF) motion control problem. The MPF motion control problem is solved for a large class of robotic vehicles that require a minimum positive forward speed to operate, which poses additional constraints, and is developed using geometric concepts, wherein the attitude control problem is formulated on Special Orthogonal group SO(3). Furthermore, the proposed control law is derived from a novel MPF error model formulation that allows to exclude the conservative constraints on the initial position of the vehicle with respect to the reference path by enabling the explicit control of the progression of a virtual point moving along the reference path. The task of the MPF control law is then to steer the vehicle towards the moving path and converge to the virtual point. \textcolor{black}{Formal stability and convergence guarantees are provided using the Input-to-State Stability concept}. In particular, we show that the proposed controller is robust to imperfect tracking errors by the autopilot and wind gusts. Simulation results are presented to illustrate the efficacy of the proposed MPF control law.
\end{abstract}
\section{Introduction}

\subsection{Motivation}

\textcolor{black}{Motion control of robotic vehicle such as Unmanned Aerial Vehicle (UAV), Autonomous Ground/Surface/Underwater Vehicle (AGV, ASV, AUV) is a fairly mature research area. The most widely researched techniques include trajectory tracking \cite{micaelli1993trajectory}, point-stabilization \cite{samson1993time} and the path following \cite{samson1992path} technique. The \textit{path following} schemes have received significant attention and requires the vehicle to follow a known geometric path at a desired nominal speed without any temporal constraints. Further, the path following scheme is known to remove performance limitations when compared with the trajectory tracking scheme \cite{aguiar2005path}. Consequently, a series of results, solely addressing the path following motion control problem were published starting with the pioneering work in \cite{micaelli1993trajectory,samson1992path,aguiar2007trajectory} for the case of wheeled mobile robots, \cite{ENCARNACAO2000507,BELLETER2016588} and references therein for the marine vehicles, and \cite{xargay2013time,kaminer2017time} for the case of UAVs.}

\textcolor{black}{Recently, a new problem termed the Moving Path Following (MPF) motion control problem was introduced \cite{tiago2013ground,tiago2016MPF}. In MPF, the robotic vehicle is tasked to follow a geometric path that is specified with respect to the reference frame that is moving. Such a problem finds application in convoy protection, target tracking, etc. The solution to the MPF problem depends on the type of the robotic vehicle and the associated motion constraints. Robotic vehicles such as unicycles, surface vessels, multi-rotors, and some AUVs can admit backward speed command (with respect to the robot's longitudinal axis) including the zero speed command. The MPF problem for such vehicles has been addressed in the authors previous work \cite{reis2019robust,jain2018moving}. These solutions however are not valid for robotic vehicles such as fixed wing UAVs and single propeller AUVs, wherein the robotic vehicles are required to have minimum positive forward speed for operation. This paper investigates the MPF motion control problem and contributes to the state-of-the-art by proposing a novel three dimensional MPF controller formulated using geometric concepts for robotic vehicles that require a minimum positive forward speed.}

\subsection{Related Work}

\textcolor{black}{This paper concerns with the MPF problem that can be viewed as a generalization of the path following motion control problem. Therefore, the literature review would be limited specifically to the relevant path following and the moving path following solutions available in the literature. As mentioned previously, the solution to the path following problem depends on the kinematic constraints of the robotic vehicle. For robots such as unicycles, wherein there are no constraints on the speed of the vehicle, i.e., the speed can be positive, negative or zero, a global Input-to-State Stable (ISS) \cite{khalil} controller is presented in \cite{aguiar2007trajectory} using the backstepping method to solve the path following problem for a generic class of underactuated vehicles in the presence of parametric modeling uncertainties. The key idea behind the proposed control strategy is to define an error variable with respect to the body frame of the robotic vehicle, that is at an offset from the the origin of the body frame (for example, the nose of the vehicle). Such a choice of error states lends itself to the path following control design with global stability properties. Similar approach was used in a recent work of \cite{zuo2018three} to design a path following controller for a stratospheric airship with assumptions that reduce the domain of attraction in order to prevent the backward movement. In the context of MPF, an approach similar to \cite{aguiar2007trajectory} was adopted for underactuated vehicles in the authors previous work \cite{jain2018cooperative}. A further extension to the ensure robustness to external disturbances, a robust MPF controller using an disturbance observer was proposed and experimentally validated in \cite{reis2019robust}. In all the above mentioned works it was assumed that the robotic vehicle is able to move forward, backward or remain stationary. Therefore, the velocity vector of the robot was governed through the reference attitude and speed commands to solve the path following objective. However, in the case of the robotic vehicles such as UAVs or certain AUVs, the speed is usually restricted to have a minimum positive forward speed. In such cases, the control objective is achieved by \textit{steering} the robotic vehicle by generating appropriate reference attitude commands, which are tracked by a lower level attitude controller.}

\textcolor{black}{To this end, early literature on the path following control utilizes a conveniently defined frame such as a Serret-Frenet frame placed at a point (referred to as `virtual point') on the given path, that is closest to the robotic vehicle \cite{samson1992path}. The control strategy is then to steer the robotic vehicle towards the virtual point by controlling the attitude while the vehicle's forward speed tracks a desired speed profile. Such a strategy, places stringent constraints on the initial position of the vehicle with respect to the path. These constraints arise out of singularities present in the kinematics of the path following system reformulated with respect to the Serret-Frenet frame. This issue was alleviated in the work of \cite{soetanto2003adaptive}, by explicitly controlling the progression of the  virtual point along the path in addition to steering the robotic vehicle to the desired path. Other studies that use the Serret-Frenet frame approach to control specific vehicles such as AUVs are presented in \cite{ENCARNACAO2000507,lapierre2007nonlinear}. A three dimensional path following controller with the attitude controller formulated in Special Orthogonal Group SO(3) was presented in \cite{cichella2011geometric} and \cite{cichella2013quadrotor} for fixed wing UAV and a multi-rotor UAV respectively. Another class of methods to solve the path following control problem, based on the missile guidance concepts are presented in \cite{breivik2009guidance,fossen2014uniform,caharija2016integral} and the references therein. Numerous other results on path following can be found in a recent survey \cite{sujit2014pathfollowing}. The path following methods based on the principles of missile guidance are most suitable for way-point following, with consecutive way-points connected through straight line segments. }

\textcolor{black}{The path following methods assume that the specified geometric path is stationary. However, in applications such a source seeking, convoy protection, target tracking, it may be useful to consider a path specified with respect to a moving frame, resulting in a Moving Path Following (MPF) motion control problem. The MPF problem was first introduced in \cite{tiago2013ground,tiago2016MPF} for an UAV tracking a ground target. Similar to the path following literature, \cite{tiago2016MPF} employs a Serret-Frenet frame approach to the 2D MPF problem. An extension to the 3D case is presented in \cite{tiago20173dMPF} where quaternions were used for the attitude representation. These approaches however assume that the virtual point to be followed is located at a point on the path that is closest to the robotic vehicle, i.e., projection of the vehicle position on the path. Consequently both these methods inherit the  problem of stringent constraints on the initial position of the robot from the path following literature. To alleviate this issue, \cite{wang2017mpf} uses technique of \cite{soetanto2003adaptive,rucco2015virtualtarget} to relax these constraints for the 2D MPF problem by explicit control of the progression of the virtual target along the path. The results however, are valid in two dimensions with the attitude of the vehicle parameterized by the heading angle. Other control methods such as vector field method \cite{KAPITANYUK20176983}, nonlinear model predictive control \cite{jain2018moving} have been proposed to solve the MPF problem for unicycle type robots. A recent work of \cite{guan2019mpf} considers a time-varying vector field guidance method to solve the MPF problem for the automated carrier landing application. A MPF control law for an ASV is presented in \cite{zheng2020mpf} where in constraints on error variables are explicitly considered through the use of barrier Lyapunov functions. All the available literature on MPF use Euler angles and quaternions for representation of attitude. While use of Euler angles result in singular configurations, the quaternions suffer from ambiguity since it double covers SO(3). A unified three dimensional MPF method that removes i) stringent constraints on the initial position of the vehicle arising out of error kinematics of the MPF system, and ii) singularities and ambiguity arising out of the chosen attitude representation is not available in the literature. This paper address precisely these issues and proposes a robust, MPF control law based on the concepts of geometric control and nonlinear control. The concepts of geometric control allows to develop control laws in a coordinate free manner.} 

%

\subsection{Contributions}

\textcolor{black}{The main contribution of the paper is the proposed solution to the 3D MPF motion control problem for robotic vehicles such as a fixed-wing UAVs or some classes of AUVs wherein the vehicles are constrained to follow a strictly positive speed profile. The constraint on the initial condition of the position of the vehicle is removed by explicitly controlling the progression of the virtual target along the moving path. Further, the concepts of geometric control theory are used to define and analyze the MPF system in order to exclude geometric singularities that arise when using local representations of attitude such as Euler angles or ambiguities when using quaternions. Specifically, assuming that the robotic vehicle tracks a desired speed profile, the error kinematics of the MPF system is obtained and the attitude control problem is formulated on Special Orthogonal group SO(3). Formal stability and convergence guarantees are provided using the Input-to-State Stability (ISS) concept along with the corresponding estimate of a region of attraction. In this paper, we adopt a practical approach based on the fact that the commercial robotic vehicles such as multi-rotors, UAVs and AUVs come with a pre-programmed autopilot control loop that is responsible to deal with the dynamics (inner-loop attitude control and control allocation) of the specific robotic vehicle. Therefore, we assume existence of such an inner-loop, stable, autopilot control loop and focus our attention towards the design of the higher level MPF guidance controller, that provides reference commands to the existing autopilot controller. In particular, we show that the MPF controller is ISS with respect to imperfect tracking autopilot errors and wind gusts. Simulation results are presented to illustrate the efficacy of the proposed MPF controller design. The theoretical result provided in this paper was used as a case study for target estimation and tracking using single range measurement in \cite{jain2018auv}. However, the stability and convergence proofs as well as the analysis in the presence of imperfect tracking by autopilot and wind disturbances were not provided - which is contained in this paper. Further, the ISS property of the controller is established considering imperfect tracking by the inner-loop autopilot controller. }

\textcolor{black}{The difference of the proposed method with respect to the existing literature can be stated as follows:
\begin{enumerate}
\item Removal of stringent constraints on the initial position of the robot found in the works of \cite{tiago2016MPF,tiago20173dMPF} by explicitly controlling the progression of the virtual point along the moving path. Further, robustness in the form of an ISS property of the proposed controller with respect to bounded error signals from the autopilot loop and wind gusts is proven. This is a stronger result compared to the works of \cite{tiago2016MPF, tiago20173dMPF} that uses Barbalat's lemma to prove stability.
\item References \cite{tiago2016MPF,zheng2020mpf} use Serret-Frenet frame that is only applicable to the paths that have non-zero curvature at all points along the curve. For paths that change the sign of the curvature such as a lemniscate path (infinity shaped path used in this paper), the curvature of the path goes to zero momentarily before switching the sign. Consequently at zero curvature, the Serret-Frenet frame is undefined. This problem is resolved in this paper through the use of Parallel Transport frame or the Bishop's frame \cite{bishop1975there}.    
\item Among the existing 3D methods of \cite{tiago20173dMPF,guan2019mpf}, this paper uses the concepts of geometric control wherein the attitude control problem is formulated in SO(3) and therefore free from singularities arising due to the use of Euler angles or ambiguities due to use of quaternions. Further, \cite{guan2019mpf} presents a time-varying vector field method to solve the MPF problem and the method proposed in this paper can be seen as an alternative method that is based on concepts of geometric control.
\item Previous works of the authors in \cite{reis2019robust,jain2018cooperative} presented a 3D MPF control law for vehicles without constraints on the speed. In this work we focus on robotic systems that require a minimum positive forward speed.
\end{enumerate}}

\textcolor{black}{The distinctive nature of the MPF control law proposed in this paper is that, it is coordinate free due to use of concepts of geometric control, free from singularities due to formulation of attitude control problem on SO(3), free from stringent constraints on the initial position of the robotic vehicle, applicable to the 3D case due to the use of the Parallel transport frame instead of Serret-Frenet frame, generic to be able to applied to any robotic system and formal robustness guarantees with respect to the autopilot tracking errors and wind disturbances - all contained in a single formulation.}

The remainder of the paper is organized as follows. Section \ref{sec:probformulation} derives the MPF position and attitude error kinematics and formulates the MPF problem addressed in this paper. The main result is presented in Section \ref{sec:mpfcontrol} that includes Lyapunov based analysis for stability of the proposed control strategy. The simulation results are illustrated in Section \ref{sec:simulationresults} followed by conclusion in Section \ref{sec:conclusion}.

\noindent \textit{Definitions and Notations --} Given a matrix $A \in \mathbb{R}^{3 \times 3}$, $A^\prime$ denotes the transpose operation, while $\lambda_{\min}(A)$ and $\lambda_{\max}(A)$ denotes the minimum and maximum eigenvalue of $A$, respectively. The configuration manifold over which the attitude of a rigid body evolves is Special Orthogonal Group $\mathrm{SO}(3)$ defined as $\mathrm{SO}(3) = \{R \in \mathbb{R}_{3 \times 3} | R^\prime R = RR^\prime = I, \hspace{0.1cm}\mathrm{det}R = I\}$, where $I = I_{3 \times 3}$ is the identity matrix unless specified otherwise. \textcolor{black}{The corresponding Lie algebra $\mathrm{so}(3)$ is the set of all $3 \times 3$ skew-symmetric matrices denoted by the \textit{hat} map $\widehat{(.)}:\mathbb{R}^3 \to \mathrm{so}(3)$.}
The inverse of the \textit{hat} map is the \textit{vee} map defined as $(.)^\vee: \mathrm{so}(3) \to \mathbb{R}^3$. 
The vectors are denoted by boldface letters and matrices are denoted by uppercase letters. The set of strictly positive real numbers are denoted by $\mathbb{R}_{>0}$. Given the notation $\boldsymbol{\omega}_{AB}^A$, the subscript denotes angular velocity $\boldsymbol{\omega}$ of the coordinate frame $\{A\}$ with respect to frame $\{B\}$, expressed in coordinate frame $\{A\}$ as denoted by superscript. Similarly, $\mathbf{x}^A$ denotes the vector $\mathbf{x}$ expressed in coordinate frame $\{A\}$ and $R_A^B$ denotes the rotation matrix from frame $\{A\}$ to frame $\{B\}$.

\section{MPF Problem Formulation}\label{sec:probformulation}
\begin{figure}
	\centering
	\includegraphics[width=0.8\linewidth]{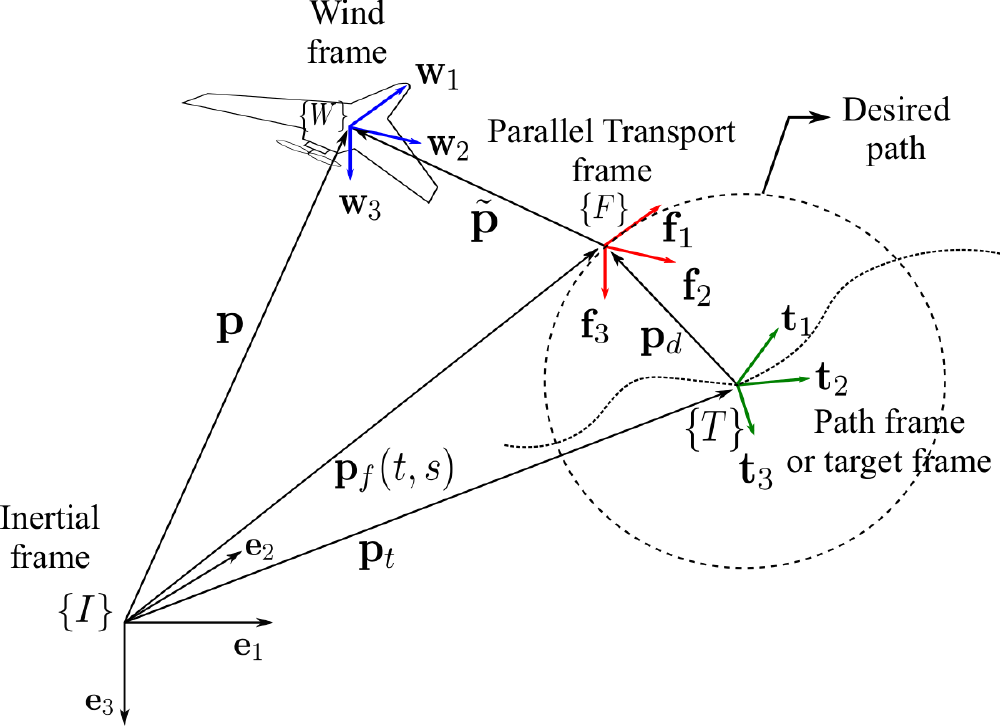}
	\caption{Coordinate frames for Moving Path Following}
	\label{fig:scheme}
\end{figure}
Consider an inertial frame of reference $\{I\} = \{{\mathbf{e}}_1,{\mathbf{e}}_2,{\mathbf{e}}_3\}$ and a Wind frame $\{W\} = \{{\mathbf{w}}_1,{\mathbf{w}}_2,{\mathbf{w}}_3\}$ with its origin attached to the center of mass of the robotic vehicle as illustrated in Figure \ref{fig:scheme}. \textcolor{black}{The wind frame, usually defined in the context of aircrafts and UAVs, is defined such that $\mathbf{w}_1$ is aligned along the velocity vector with respect to the fluid of the robotic vehicle. For UAVs, see e.g. \cite{beard2012small}, the $\mathbf{w}_1$ axis points along the airspeed vector, which is equivalent to ground speed vector in the absence of wind}. Let the position of the vehicle be denoted by $\mathbf{p}(t) \in \mathbb{R}^3$ and its attitude denoted by the rotation matrix $R_W^I \in \mathrm{SO}(3)$. Assuming that the position and the attitude of the vehicle is known or measured, the kinematic model satisfies,
\begin{align}\label{eq:UAVKinematics}
\dot{\mathbf{p}} &= v_w\mathbf{w}_1 \\
\dot{R}_W^I &= R_W^I\widehat{\boldsymbol{\omega}}_{WI}^W
\end{align}
where $v_w \in \mathbb{R}_{>0}$ is the speed of the vehicle acting along the longitudinal direction $\mathbf{w}_1$ that can be considered constant without loss of generality and constrained as
\begin{equation}
0 < v_{w,\mathrm{min}} \leq v_w \leq v_{w,\mathrm{max}}
\end{equation}
Such a restriction arises for fixed-wing UAVs and certain classes of AUVs that cannot have a zero speed \textcolor{black}{with respect to the surrounding fluid} during their operation. Further, define a Path Transport frame or a target frame $\{T\} = \{{\mathbf{t}}_1,{\mathbf{t}}_2,{\mathbf{t}}_3\}$, fixed to a known moving target with position $\mathbf{p}_t \in \mathbb{R}^3$, linear velocity $\dot{\mathbf{p}}_t \in \mathbb{R}^3$, attitude $R_T^I \in \mathrm{SO}(3)$ and angular velocity $\boldsymbol{\omega}_{TI} \in \mathbb{R}^3$ that are assumed to be known a priori.
Let $\mathbf{p}_d:\mathbb{R} \to \mathbb{R}^3$ denote the reference geometric path, parameterized by $s \in \mathbb{R}$, that needs to be followed by the vehicle and \textit{is stationary} with respect to the Path Transport frame $\{T\}$. The parameter $s$ is the arc length along the reference path and for a given $s$, $\mathbf{p}_d(s)$ denotes a virtual point on the reference path. Define a Parallel Transport Frame (also referred to as Bishop's frame \cite{bishop1975there}) $\{F\} = \{{\mathbf{f}}_1,{\mathbf{f}}_2,{\mathbf{f}}_3\}$ attached to this point.
The rotation matrix from frame $\{F\}$ to the frame $\{T\}$ is denoted by $R_F^T = [\mathbf{f}_1^{T} \hspace{0.1cm} \mathbf{f}_2^{T} \hspace{0.1cm} \mathbf{f}_3^{T}]$ that can be computed from the given path as explained in \cite{hanson1995parallel}.
\begin{remark}
	The advantage of using the Parallel Transport Frame is that it is well defined when the path has a vanishing second derivative. This is in contrast with the Serret-Frenet frame that is undefined when the path curvature vanishes for example in straight line segments. \qed
\end{remark}
\subsection{MPF Position Error Kinematics}
Let $\tilde{\mathbf{p}}$ denote the MPF position error between the position of the robotic vehicle and the origin of the parallel transport frame $\{F\}$. Then, $\tilde{\mathbf{p}} = \mathbf{p} - \mathbf{p}_f$, where $\mathbf{p}_f = \mathbf{p}_t + \mathbf{p}_d(s)$ is the desired position of the vehicle. The time derivative of the desired position vector $\mathbf{p}_f$ is given as,
\begin{equation}
\dot{\mathbf{p}}_f = v_t\mathbf{t}_1 + \boldsymbol{\omega}_{TI} \times \mathbf{p}_d + \dot{s}\mathbf{f}_1 
\end{equation}
where $v_t = \|\dot{\mathbf{p}}_t\|$, $\mathbf{t}_1 = \frac{\dot{\mathbf{p}}_t}{\|\dot{\mathbf{p}}_t\|}$ and $v_t\mathbf{t}_1 + \boldsymbol{\omega}_{TI} \times \mathbf{p}_d$ represents the velocity of the desired position along the path due to the motion of the target frame $\{T\}$. The MPF position error kinematics is given by
\begin{equation} \label{eq:MPFpositionerrorkinematics}
\dot{\tilde{\mathbf{p}}} = v_w{\mathbf{w}}_1 - \dot{s}{\mathbf{f}}_1 - {v}_t{\mathbf{t}}_1 - \boldsymbol{\omega}_{{TI}} \times \mathbf{p}_d -  \boldsymbol{\omega}_{{FT}} \times \tilde{\mathbf{p}}
\end{equation}
where $\boldsymbol{\omega}_{{FT}} \times \tilde{\mathbf{p}}$ represents the contribution  of the rotational motion of the parallel transport frame $\{F\}$.
\textcolor{black}{\begin{remark}
		The existing literature on MPF \cite{tiago2016MPF,tiago20173dMPF} requires computation of the virtual point on the path that is closest to the vehicle, which is equivalent to the projection of the robot position onto the moving path. This places stringent constraints on the initial position of the vehicle. In this paper, these constraints are eliminated by explicitly controlling the progression of the virtual point along the path, thereby treating $\dot{s}$ as a virtual control input to the system. \qed
\end{remark}}
\subsection{MPF Attitude Error Kinematics}
Given that the robotic vehicle is assumed to have a constant speed $v_w$, the MPF motion control problem is solved by shaping its attitude towards the moving path, aided by the the virtual control input $\dot{s}$ that enables faster convergence. To this end, two additional coordinate frames are introduced, namely $\{W_d\} = \{\mathbf{w_d}_1, \mathbf{w_d}_2, \mathbf{w_d}_3\}$ and $\{D\} = \{\mathbf{d}_1, \mathbf{d}_2, \mathbf{d}_3\}$ with rotation matrices $R_{W_d}^F$ and $R_D^{W_d}$ respectively. The basis vector $\mathbf{w_d}_1$ defines the steady state desired direction of the velocity vector of the vehicle when it converges to the moving path. The basis vector $\mathbf{d}_1$ defines the desired direction of the velocity vector of the vehicle during the transient phase. Therefore the basis vector $\mathbf{d}_1$ must be defined such that it smoothly converges to the steady state vector $\mathbf{w_d}_1$, thereby shaping the approach attitude of the vehicle to the moving path. Consequently, the objective of the attitude controller is to ensure that the direction of the velocity vector of the vehicle denoted by $\mathbf{w}_1$ aligns with the desired direction $\mathbf{d}_1$, i.e., $\mathbf{w}_1\cdot\mathbf{d}_1 = 1$, by controlling the angular velocities $\boldsymbol{\omega}_{WI}^W$ of the vehicle.

Therefore, define a real-valued error function $\Psi:\mathrm{SO}(3) \to \mathbb{R}$ \cite{xargay2013time} as,
\begin{equation}\label{eq:attitudeErrorFunction}
\Psi(\tilde{R}) = \frac{1}{2}\mathrm{tr}\left[(I_3 - \Pi_R^\prime\Pi_R)(I_3 - \tilde{R})\right] = \frac{1}{2}\left(1 - \tilde{R}_{11}\right)
\end{equation}
where $\tilde{R} = R_W^D$ is the rotation matrix that denotes the attitude error from $\{W\}$ frame to the $\{D\}$ frame and $\Pi_R = \left[\begin{array}{ccc}
0 & 1 & 0 \\ 
0 & 0 & 1
\end{array} \right]$. 
The matrix $\Pi_R$ is the selector matrix, that selects the $(1,1)$ entry of the rotation matrix $\tilde{R}$ denoted as $\tilde{R}_{11}$. Notice that $\tilde{R}_{11}$ is equivalent to $\mathbf{w}_1\cdot\mathbf{d}_1$ and hence, the attitude error function is positive definite about $\tilde{R}_{11} = \mathbf{w}_1\cdot\mathbf{d}_1 = 1$. Following similar steps described in \cite{xargay2013time}, the MPF attitude error kinematics is obtained by differentiating the attitude error function \eqref{eq:attitudeErrorFunction} with respect to time and satisfies
\begin{equation}\label{eq:attitudeKinematics}
\dot{\Psi}(\tilde{R}) = \mathbf{e}_{\tilde{R}}\cdot\Pi_R\boldsymbol{\omega}_{WD}^W
\end{equation}
where 
\begin{align}
\mathbf{e}_{\tilde{R}} &= \frac{1}{2}\Pi_R\left((I_3 - \Pi_R^\prime\Pi_R)\tilde{R} - \tilde{R}^\prime(I_3 - \Pi_R^\prime\Pi_R)\right)^\vee \\ 
&= \frac{1}{2}\left[\tilde{R}_{13} \hspace{0.2cm} {-\tilde{R}_{12}}\right]^\prime \label{eq:ertilde}
\end{align}
defines the attitude error vector. Note that $\|\mathbf{e}_{\tilde{R}}\| \to 0$ implies, $\tilde{R}_{11} \to 1$ and consequently $\Psi(\tilde{R}) \to 0$. The term $\Pi_R\boldsymbol{\omega}_{WD}^W$ satisfies
\begin{equation}\label{eq:ffomega}
\Pi_R\boldsymbol{\omega}_{WD}^W = \Pi_R\boldsymbol{\omega}_{WI}^W - \Pi_R\left(\boldsymbol{\omega}_{TI}^W + \boldsymbol{\omega}_{FT}^W + \boldsymbol{\omega}_{{W_d}F}^W  + \boldsymbol{\omega}_{D{W_d}}^W\right)
\end{equation}
that can be written in terms of known quantities as
\begin{multline}
\Pi_R\boldsymbol{\omega}_{WD}^W =  \Pi_R\boldsymbol{\omega}_{WI}^W - \Pi_R\tilde{R}^\prime\left(R_T^D\boldsymbol{\omega}_{TI}^T\right.\\
\left. + R_F^D\boldsymbol{\omega}_{FT}^F + R_{W_d}^D\boldsymbol{\omega}_{{W_d}D}^{W_d} + \boldsymbol{\omega}_{D{W_d}}^D\right)
\end{multline}
\textcolor{black}{Equations \eqref{eq:MPFpositionerrorkinematics} and \eqref{eq:attitudeKinematics} represent the MPF position error and attitude error kinematics respectively. The term $\Pi_R\boldsymbol{\omega}_{WI}^W$ implies that the direction of the velocity vector of the vehicle $\mathbf{w}_1$ is dependent on the angular velocities about the $\mathbf{w}_2$-$\mathbf{w}_3$ axes. Introduce a signal $\bar{\boldsymbol{\omega}}_{WI}^W = [0 \hspace{0.1cm} (\Pi_R\bar{\boldsymbol{\omega}}_{WI}^W)^\prime]^\prime$ that forms the control references to an existing inner-loop autopilot controller. The objective of the autopilot is to ensure that the angular velocity of the wind frame attached to the vehicle ${\boldsymbol{\omega}}_{WI}^W = [0 \hspace{0.1cm} (\Pi_R{\boldsymbol{\omega}}_{WI}^W)^\prime]^\prime$ tracks the control reference signal $\bar{\boldsymbol{\omega}}_{WI}^W$. To this end, consider the autopilot tracking error signal $\tilde{\boldsymbol{\omega}} := \Pi_R{\boldsymbol{\omega}}_{WI}^W - \Pi_R\bar{\boldsymbol{\omega}}_{WI}^W$ that is assumed to be bounded.} The MPF motion control problem is stated as follows:
\begin{problem}[Moving Path Following]
\textcolor{black}{Given a desired geometric, regular path $\mathbf{p}_d(s)$ that is stationary with respect to the target frame $\{T\}$, the quantities $\mathbf{p}_t(t)$, $\dot{\mathbf{p}}_t(t)$, $R_T^I$, and $\boldsymbol{\omega}_{TI}^T$ related to the target motion, and an existing autopilot controller such that $\tilde{\boldsymbol{\omega}}$ is bounded, the MPF motion control problem here is to design control laws for the virtual control input $\dot{s}$ and the reference signal for the autopilot $\bar{\boldsymbol{\omega}}_{WI}^W$, such that the MPF position error $\|\tilde{\mathbf{p}}\|$ and the MPF attitude error $\|\mathbf{e}_{\tilde{R}}\|$ converges to a small neighborhood of zero as $t \to \infty$. Further, in the absence of $\tilde{\boldsymbol{\omega}}$, the signals $\|\tilde{\mathbf{p}}\|$ and $\|\mathbf{e}_{\tilde{R}}\|$ should converge to zero.} \qed
\end{problem}
\begin{remark}
	Note that when the target is stationary, i.e., $\dot{\mathbf{p}}_t = 0$ and $\boldsymbol{\omega}_{TI} = 0$, the position error and attitude error kinematics of \eqref{eq:MPFpositionerrorkinematics} and \eqref{eq:attitudeKinematics} respectively, reduce to the classical path following problem with $R_{W_d}^F = I_3$. 
\end{remark}

\section{3D Moving Path Following} \label{sec:mpfcontrol}
Consider the following assumptions on the MPF system required for a well-posed MPF problem.
\begin{assume}\label{ass:one}
The desired speed of the robotic vehicle $v_w$ satisfies $v_w > \big \|v_t\mathbf{t}_1 + \boldsymbol{\omega}_{TI} \times \mathbf{p}_d \big \|$ for all time $t \in \mathbb{R}$. The condition implies that the vehicle has sufficient speed to catch up to the target motion.
\end{assume}
The rotation matrix $R_{W_d}^F$ that defines the desired steady state attitude of the vehicle velocity vector is a function of the velocities of the target frame according  to the next definition.
\begin{definition}[Desired steady state attitude] The desired steady-state coordinate frame $\{W_d\}$ is defined by the orthonormal basis vectors $\mathbf{w_d}_1$, $\mathbf{w_d}_2$ and $\mathbf{w_d}_3$ given by,
\begin{align}\label{eq:Wd_basis}
\mathbf{w_d}_1 &:= {w_d}_{11}\mathbf{f}_1 + {w_d}_{21}\mathbf{f}_2 + {w_d}_{31}\mathbf{f}_3 \\
\mathbf{w_d}_2 &:= {w_d}_{12}\mathbf{f}_1 + {w_d}_{22}\mathbf{f}_2 + {w_d}_{32}\mathbf{f}_3 \\
\mathbf{w_d}_3 &:= \mathbf{w_d}_1 \times \mathbf{w_d}_2  = {w_d}_{13}\mathbf{f}_1 + {w_d}_{23}\mathbf{f}_2 + {w_d}_{33}\mathbf{f}_3
\end{align}
where
\begin{equation}\label{eq:wdbasis}
\begin{array}{ll}
{w_d}_{11} = \sqrt{1 - {w_d}_{21}^2 - {w_d}_{31}^2 } & {w_d}_{12} = \frac{-{w_d}_{21}}{\sqrt{{w_d}_{11}^2 + {w_d}_{21}^2}}  \\ 
{w_d}_{21} = \frac{v_t}{v_w}\mathbf{t}_1\cdot\mathbf{f}_2 + \frac{(\boldsymbol{\omega}_{TI} \times \mathbf{p}_d)\cdot\mathbf{f}_2}{v_w} & {w_d}_{22} = \frac{{w_d}_{11}}{\sqrt{{w_d}_{11}^2 + {w_d}_{21}^2}} \\ 
{w_d}_{31} = \frac{v_t}{v_w}\mathbf{t}_1\cdot\mathbf{f}_3 + \frac{(\boldsymbol{\omega}_{TI} \times \mathbf{p}_d)\cdot\mathbf{f}_3}{v_w} & {w_d}_{32} = 0
\end{array}
\end{equation} 
\end{definition}
\begin{property}The coordinate frame defined in \eqref{eq:Wd_basis} - \eqref{eq:wdbasis} with the rotation matrix $R_{W_d}^F = \left[\mathbf{w_d}_1 \hspace{0.2cm} \mathbf{w_d}_2 \hspace{0.2cm} \mathbf{w_d}_3\right]$
has the same steady state attitude of the velocity vector of the robotic vehicle as required to solve the MPF problem. More precisely, $\mathbf{w_d}_1\cdot\bar{\mathbf{w}}_1 = 1$, where $\bar{\mathbf{w}}_1$ is the velocity vector of robotic vehicle in steady state.
\begin{proof}
During the steady state, the MPF position error $\tilde{\mathbf{p}} = 0$ and $\dot{\tilde{\mathbf{p}}} = 0$. Thus, under these conditions the MPF position error kinematics described in \eqref{eq:MPFpositionerrorkinematics} yields
\begin{equation}\label{eq:property_equation}
\bar{\mathbf{w}}_1 = \frac{\dot{s}}{v_w}\mathbf{f}_1 + \frac{v_t}{v_w}\mathbf{t}_1 + \frac{\boldsymbol{\omega}_{TI} \times \mathbf{p}_d}{v_w}
\end{equation}
Therefore, computing the projection of vector $\bar{\mathbf{w}}_1$ on the axes of $\{F\}$ frame results in $\mathbf{w_d}_1$ with its components along $\mathbf{f}_2$ and $\mathbf{f}_3$ given by ${w_d}_{21} = \bar{\mathbf{w}}_1\cdot\mathbf{f}_2 = \frac{v_t}{v_w}\mathbf{t}_1\cdot\mathbf{f}_2 + \frac{(\boldsymbol{\omega}_{TI} \times \mathbf{p}_d)\cdot\mathbf{f}_2}{v_w}$, and ${w_d}_{31} = \bar{\mathbf{w}}_1\cdot\mathbf{f}_3 = \frac{v_t}{v_w}\mathbf{t}_1\cdot\mathbf{f}_3 + \frac{(\boldsymbol{\omega}_{TI} \times \mathbf{p}_d)\cdot\mathbf{f}_3}{v_w}$, respectively. Note that from Assumption \ref{ass:one}, ${w_d}_{21} < 1$ and ${w_d}_{31} < 1$ is always true and since $ {w_d}_{11} = \sqrt{1 - {w_d}_{21}^2 - {w_d}_{31}^2 }$, then ${w_d}_{11}^2 + {w_d}_{21}^2 + {w_d}_{31}^2 = 1$. Choosing ${w_d}_{12}$, ${w_d}_{22}$ and ${w_d}_{32}$ as given in \eqref{eq:wdbasis} ensures that $\mathbf{w_d}_1\cdot\mathbf{w_d}_2 = 0$. The computation of $\mathbf{w_d}_3$ follows from taking the cross product of the basis vectors $\mathbf{w_d}_1$ and $\mathbf{w_d}_2$.
\end{proof}
\end{property}
The task of the rotation matrix $R_D^{W_d}$ is to shape the attitude of the vehicle such that the velocity vector $\mathbf{w}_1$ converges smoothly to the desired velocity vector $\mathbf{w_d}_1$. Intuitively, when the vehicle is far away from the path, the desired velocity vector $\mathbf{d}_1$ is almost perpendicular to the desired steady state vector $\mathbf{w_d}_1$. However, as the MPF position error becomes smaller in magnitude, the vector $\mathbf{d}_1$ should align to the vector $\mathbf{w_d}_1$. To this end, consider the following cross track position error vector defined as
\begin{equation}
\tilde{\mathbf{p}}_\times := (\tilde{\mathbf{p}}\cdot\mathbf{f}_2)\mathbf{f}_2 + (\tilde{\mathbf{p}}\cdot\mathbf{f}_3)\mathbf{f}_3
\end{equation}
Further, consider the projection of vector $\tilde{\mathbf{p}}_\times$ onto the plane spanned by the basis vectors $\mathbf{w_d}_2 - \mathbf{w_d}_3$ defined as
\begin{align}
\breve{\mathbf{p}}_\times &:= (\tilde{\mathbf{p}}_\times\cdot\mathbf{w_d}_2)\mathbf{w_d}_2 + (\tilde{\mathbf{p}}_\times\cdot\mathbf{w_d}_3)\mathbf{w_d}_3 \\
&:= y_w\mathbf{w_d}_2 + z_w\mathbf{w_d}_3
\end{align}
The rotation matrix $R_D^{W_d}$ that defines the desired transient attitude is given as
\begin{equation}
R_D^{W_d} = \left[\mathbf{d}_1 \hspace{0.1cm} \mathbf{d}_2 \hspace{0.1cm} \mathbf{d}_3\right] = \left[\begin{array}{ccc}
{d}_{11} & {d}_{12} & {d}_{13} \\ 
{d}_{21} & {d}_{22} & {d}_{23} \\ 
{d}_{31} & {d}_{32} & {d}_{33}
\end{array} \right]
\end{equation}
where
\begin{equation*}
\begin{array}{ll}
{d}_{11} = \frac{\alpha}{\sqrt{\alpha^2 + y_w^2 + z_w^2}} & {d}_{12} = \frac{y_w}{\sqrt{\alpha^2 + y_w^2}}  \\ 
{d}_{21} = \frac{-y_w}{\sqrt{\alpha^2 + y_w^2 + z_w^2}} & {d}_{22} =\frac{\alpha}{\sqrt{\alpha^2 + y_w^2}} \\ 
{d}_{31} = \frac{-z_w}{\sqrt{\alpha^2 + y_w^2 + z_w^2}} & {d}_{32} = 0
\end{array}
\end{equation*} 
for any $\alpha > 0$ and $\mathbf{d}_3 := \mathbf{d}_1 \times \mathbf{d}_2$. The MPF control law is then given by the following theorem.
\begin{theorem}[Moving Path Following Control]
Consider the MPF position error and attitude error kinematics given by \eqref{eq:MPFpositionerrorkinematics} and \eqref{eq:attitudeKinematics} respectively, along with Assumption \ref{ass:one}. Let $c_1 > 0$ be a constant that can be made arbitrarily large. Let the initial conditions satisfy
	\begin{equation}\label{eq:DOAInitPositionAndAttitude}
	\|\tilde{\mathbf{p}}(0)\| \leq c_1 \mbox{ and } \Psi(\tilde{R}(0)) \leq c_2 < \frac{1}{2}
	\end{equation}
Then, the virtual control input $\dot{s}$ given by
\begin{equation}\label{eq:sdot}
\dot{s} = \left(K_p\tilde{\mathbf{p}} + v_w\mathbf{w}_1 - v_t\mathbf{t}_1 - \boldsymbol{\omega}_{TI} \times \mathbf{p}_d\right)\cdot\mathbf{f}_1
\end{equation}
along with the reference signal to the autopilot
\begin{multline}\label{eq:omega}
\Pi_R\bar{\boldsymbol{\omega}}_{WI}^W = -K_{\tilde{R}}\mathbf{e}_{\tilde{R}} + \Pi_R\tilde{R}^\prime\left(R_T^D\boldsymbol{\omega}_{TI}^T\right.\\
\left. + R_F^D\boldsymbol{\omega}_{FT}^F + R_{W_d}^D\boldsymbol{\omega}_{{W_d}D}^{W_d} + \boldsymbol{\omega}_{D{W_d}}^D\right)
\end{multline}
where $K_p > 0$ and $K_{\tilde{R}} > 0$ are controller gains, makes the overall MPF system Input-to-State Stable (ISS) with respect to the autopilot tracking error signal $\tilde{\boldsymbol{\omega}}$ within the region of attraction,
\begin{equation}\label{eq:DOA}
\Omega_{\mathrm{MPF}} := \left\lbrace(\tilde{\mathbf{p}}, \tilde{R}) \in \mathbb{R}^3 \times \mathrm{SO}(3) | \Psi(\tilde{R}) + \frac{1}{4c_1^2}\|\tilde{\mathbf{p}}\|^2 \leq \bar{c}\right\rbrace
\end{equation}
where $\bar{c} \in \left(0, \frac{1}{2}\right)$, provided that the controller gains satisfy
\begin{equation}\label{eq:conditionsGainAndK}
KK_{\tilde{R}} > \frac{v_{w, \max}^2}{c_1^2(1-2c_2)^2} \mbox{ with } K := \min\left\lbrace K_p, \frac{v_{w,\min}}{(\alpha^2 + c_1^2)^{\frac{1}{2}}}\right\rbrace
\end{equation}
for any $\alpha > 0$.
\begin{proof}
\textit{Step 1: Boundedness of $\Psi(\tilde{R})$ --} The attitude error function satisfies $\Psi(\tilde{R}) = \frac{1}{2}(1 - \tilde{R}_{11})$ and therefore $\tilde{R}_{11} = 1 - 2\Psi$. Additionally, from the definition of the attitude error vector $\mathbf{e}_{\tilde{R}}$ in \eqref{eq:ertilde} we have, $\|\mathbf{e}_{\tilde{R}}\|^2 = \frac{1}{4}(\tilde{R}_{12}^2 + \tilde{R}_{13}^2)$. Using the relation $\tilde{R}_{11}^2 + \tilde{R}_{12}^2 + \tilde{R}_{13}^2 = 1$, it follows that $\|\mathbf{e}_{\tilde{R}}\|^2 = \Psi (1 - \Psi)$. 
The attitude error function $\Psi(\tilde{R})$ is bounded as $0 \leq \Psi(\tilde{R}) \leq c_2 < 1$, where the bound on $c_2$ is yet to be determined. Using these bounds, the attitude error function can be bounded as
\begin{equation}\label{eq:boundPsi}
\|\mathbf{e}_{\tilde{R}}\|^2 \leq \Psi(\tilde{R}) \leq \frac{1}{1 - c_2}\|\mathbf{e}_{\tilde{R}}\|^2
\end{equation} 
\textit{Step 2: Candidate ISS Lyapunov Function --} Consider the ISS Lyapunov function candidate 
\begin{equation}
V(x_\mathrm{MPF}) := k_1\|\tilde{\mathbf{p}}\|^2 + \frac{\|\mathbf{e}_{\tilde{R}}\|^2}{1 - \Psi(\tilde{R})} = k_1\tilde{\mathbf{p}}\cdot\tilde{\mathbf{p}} + \Psi(\tilde{R})
\end{equation}
where $x_\mathrm{MPF} = \left[\|\tilde{\mathbf{p}}\| \hspace{0.2cm} \|\mathbf{e}_{\tilde{R}}\|\right]^\prime$ and $V(x_\mathrm{MPF})$ is positive definite as $\Psi(\tilde{R}) \leq c_2 < 1$. 
Therefore the candidate ISS Lyapunov function can be bounded using \eqref{eq:boundPsi} as 
\begin{equation}\label{eq:boundLyapFunc}
x_\mathrm{MPF}^\prime M_1 x_\mathrm{MPF} \leq V(x_\mathrm{MPF}) \leq x_\mathrm{MPF}^\prime M_2 x_\mathrm{MPF}
\end{equation}
with  $M_1 = \mathrm{diag}(k_1, 1)$ and $M_2 = \mathrm{diag}(k_1, \frac{1}{1 - c_2})$.\\
\noindent \textit{Step 3: Derivative of ISS Lyapunov Function --} Taking the time derivative, using \eqref{eq:attitudeKinematics} and \eqref{eq:omega}, with $\Pi_R\boldsymbol{\omega}_{WI}^W = \Pi_R\bar{\boldsymbol{\omega}}_{WI}^W + \tilde{\boldsymbol{\omega}}$
\begin{equation}
\dot{V} = 2k_1\tilde{\mathbf{p}}\cdot\dot{\tilde{\mathbf{p}}} -K_{\tilde{R}} \mathbf{e}_{\tilde{R}}\cdot\mathbf{e}_{\tilde{R}} + \mathbf{e}_{\tilde{R}} \cdot\tilde{\boldsymbol{\omega}}
\end{equation}
Using equation \eqref{eq:MPFpositionerrorkinematics}, \eqref{eq:sdot} and $\tilde{\mathbf{p}}\cdot\left(\boldsymbol{\omega}_{{FT}} \times \tilde{\mathbf{p}}\right) = 0$, the term $\tilde{\mathbf{p}}\cdot\dot{\tilde{\mathbf{p}}}$ can be written as
\begin{equation*}
\tilde{\mathbf{p}}\cdot\dot{\tilde{\mathbf{p}}} = -K_p\left(\tilde{\mathbf{p}}\cdot\mathbf{f}_1\right)^2 + (v_w\mathbf{w}_1 - v_t\mathbf{t}_1 - \mathbf{v_d})\cdot\tilde{\mathbf{p}}_\times
\end{equation*}
where $\mathbf{v_d} = \boldsymbol{\omega}_{TI} \times \mathbf{p}_d$, and $\tilde{\mathbf{p}}_\times = \left[\tilde{\mathbf{p}} - (\tilde{\mathbf{p}}\cdot\mathbf{f}_1)\mathbf{f}_1\right]$. Substituting in $\dot{V}$, we have
\begin{multline*}
\dot{V} = -K_{\tilde{R}}\|\mathbf{e}_{\tilde{R}}\|^2 - 2k_1K_p\left(\tilde{\mathbf{p}}\cdot\mathbf{f}_1\right)^2\\ + 2k_1v_w\left(\mathbf{w}_1 - \frac{v_t}{v_w}\mathbf{t}_1 -\frac{\mathbf{v_d}}{v_w}\right)\cdot\tilde{\mathbf{p}}_\times + \mathbf{e}_{\tilde{R}} \cdot\tilde{\boldsymbol{\omega}}
\end{multline*}
From the fact that $\left(\frac{v_t}{v_w}\mathbf{t}_1 + \frac{\mathbf{v_d}}{v_w}\right)\cdot\tilde{\mathbf{p}}_\times = \tilde{\mathbf{p}}_\times \cdot \mathbf{w_d}_1$ it follows that
\begin{multline*}
\dot{V} = -K_{\tilde{R}}\|\mathbf{e}_{\tilde{R}}\|^2 - 2k_1K_p\left(\tilde{\mathbf{p}}\cdot\mathbf{f}_1\right)^2\\ + 2k_1v_w\tilde{\mathbf{p}}_\times\cdot(\mathbf{w}_1 - \mathbf{w_d}_1) + \mathbf{e}_{\tilde{R}} \cdot\tilde{\boldsymbol{\omega}}
\end{multline*}
From the definition of $R_D^{W_d}$, we have $0 < \frac{\alpha}{\sqrt{\alpha^2 + y_w^2 + z_w^2}} \leq \mathbf{d}_1\cdot\mathbf{w_d}_1 \leq 1$. 
Further, note that $\mathbf{w}_1\cdot\mathbf{d}_1 = \tilde{R}_{11} = 1 - 2\Psi(\tilde{R})$. Since, $0 \leq \Psi(\tilde{R}) \leq c_2$, the bound $1- 2c_2 \leq \mathbf{w}_1\cdot\mathbf{d}_1 \leq 1$ holds.
Therefore, $2k_1v_w\tilde{\mathbf{p}}_\times\cdot(\mathbf{w}_1 - \mathbf{w_d}_1)$ can be upper bounded as
\begin{align}
 &\leq \frac{2k_1v_w}{\mathbf{w}_1\cdot\mathbf{d}_1}\tilde{\mathbf{p}}_\times\cdot \left(\mathbf{w}_1 - (\mathbf{w}_1\cdot\mathbf{d}_1)\mathbf{d}_1 \right) + 2k_1v_w\tilde{\mathbf{p}}_\times\cdot\mathbf{d}_1^\times \nonumber \\
 &\leq \frac{2k_1v_w}{\mathbf{w}_1\cdot\mathbf{d}_1}\tilde{\mathbf{p}}_\times\cdot \left(\mathbf{d}_1 \times (\mathbf{w}_1\times\mathbf{d}_1)\right) + 2k_1v_w\tilde{\mathbf{p}}_\times\cdot\mathbf{d}_1^\times \label{eq:I1}
\end{align}
where $\mathbf{d}_1^\times = (\mathbf{d}_1\cdot\mathbf{w_d}_2)\mathbf{w_d}_2 + (\mathbf{d}_1\cdot\mathbf{w_d}_3)\mathbf{w_d}_3$. The term $\mathbf{w}_1\cdot\mathbf{d}_1 > 0$ when $c_2 < \frac{1}{2}$ leading to the condition \eqref{eq:DOAInitPositionAndAttitude}. Further, $\|\breve{\mathbf{p}}\| \leq \|\tilde{\mathbf{p}}_\times\| \leq \|\tilde{\mathbf{p}}\| \leq c_1$ holds due to the condition \eqref{eq:DOAInitPositionAndAttitude}. Therefore, $\tilde{\mathbf{p}}_\times\cdot\mathbf{d}_1^\times$ can be written as
\begin{equation}\label{eq:I2}
\tilde{\mathbf{p}}_\times\cdot\mathbf{d}_1^\times = \frac{-\breve{\mathbf{p}}_\times\cdot\breve{\mathbf{p}}_\times}{\sqrt{\alpha^2 + \breve{\mathbf{p}}_\times\cdot\breve{\mathbf{p}}_\times}} \leq \frac{-\tilde{\mathbf{p}}_\times\cdot\tilde{\mathbf{p}}_\times}{\sqrt{\alpha^2 + c_1^2}}
\end{equation}
Substituting \eqref{eq:I1} and \eqref{eq:I2} in the time derivative of the candidate Lyapunov function,
\begin{multline*}
\dot{V} \leq-K_{\tilde{R}}\|\mathbf{e}_{\tilde{R}}\|^2 -2k_1K\|\tilde{\mathbf{p}}\|^2 \\+  \frac{2k_1v_{w,\mathrm{max}}}{1 - 2c_2}\|\tilde{\mathbf{p}}_\times\|\|\left(\mathbf{d}_1 \times (\mathbf{w}_1\times\mathbf{d}_1)\right)\| + \|\mathbf{e}_{\tilde{R}}\|\| \tilde{\boldsymbol{\omega}}\|
\end{multline*}
where $K$ is given by \eqref{eq:conditionsGainAndK} (See \cite{xargay2013time}). Further, $\|\left(\mathbf{d}_1 \times (\mathbf{w}_1\times\mathbf{d}_1)\right)\|$ represents the magnitude of the sine of angle between the vectors $\mathbf{d}_1$ and $\mathbf{w}_1$. From \cite{xargay2013time}, it can be verified that $\|\left(\mathbf{d}_1 \times (\mathbf{w}_1\times\mathbf{d}_1)\right)\| = 2\|\mathbf{e}_{\tilde{R}}\|$. Further using young's inequality $\|\mathbf{e}_{\tilde{R}}\|\| \tilde{\boldsymbol{\omega}}\| \leq (a/2)\|\mathbf{e}_{\tilde{R}}\|^2 + (1/2a)\| \tilde{\boldsymbol{\omega}}\|^2$ with $a = K_{\tilde{R}}$, we have 
	\begin{equation}\label{eq:ISS_decay}
	\dot{V} \leq -x_\mathrm{MPF}^\prime Wx_\mathrm{MPF} + \frac{1}{2K_{\tilde{R}}}\| \tilde{\boldsymbol{\omega}}\|^2 \leq - \lambda_{\mathrm{MPF}}V  + \frac{1}{2K_{\tilde{R}}}\| \tilde{\boldsymbol{\omega}}\|^2
	\end{equation}
	where 
	\[\lambda_{\mathrm{MPF}} = \frac{\lambda_{\min}(W)}{\lambda_{\max}(M_2)} \mbox{ and } W = \left[\begin{array}{cc}
	2k_1K&  -\frac{2k_1v_{w,\mathrm{max}}}{1 - 2c_2}\\ 
	-\frac{2k_1v_{w,\mathrm{max}}}{1 - 2c_2} & \frac{K_{\tilde{R}}}{2}
	\end{array} \right]\] The condition $\lambda_{\mathrm{MPF}} > 0$ implies $W > 0$, that holds true provided the condition \eqref{eq:conditionsGainAndK} is satisfied. We can now conclude that the MPF system \eqref{eq:MPFpositionerrorkinematics} and \eqref{eq:attitudeKinematics} with the proposed control law \eqref{eq:sdot} and \eqref{eq:omega} is ISS with respect to the autopilot tracking error signal $\tilde{\boldsymbol{\omega}}$. A conservative estimate of the domain of attraction is given by $\Omega_{\mathrm{MPF}} := \left\lbrace(\tilde{\mathbf{p}}, \tilde{R}) \in \mathbb{R}^3 \times \mathrm{SO}(3) | V(x_\mathrm{MPF}) \leq \bar{c} < c = \frac{1}{2}\right\rbrace$ where $c$ is obtained using the bounds \eqref{eq:DOAInitPositionAndAttitude} and \eqref{eq:boundLyapFunc} as $c = \min_{\Psi = \frac{1}{2}, \|\tilde{\mathbf{p}}\| = c_1}V(x_\mathrm{MPF}) = \frac{1}{4} + k_1 c_1^2 $. Set $k_1 = \frac{1}{4c_1^2}$ to obtain \eqref{eq:DOA}. The solution to \eqref{eq:ISS_decay} can be written as $V(t) \leq e^{-\lambda_{\mathrm{MPF}}t}V(0) + \frac{\epsilon}{2K_{\tilde{R}}\lambda_{\mathrm{MPF}}} \leq \bar{c} + \frac{\epsilon}{2K_{\tilde{R}}\lambda_{\mathrm{MPF}}} < c$.
\end{proof}
\end{theorem}
\begin{figure*}[h]
	\begin{subfigure}[t]{.5\textwidth}
		\centering
		\includegraphics[width=1.0\linewidth]{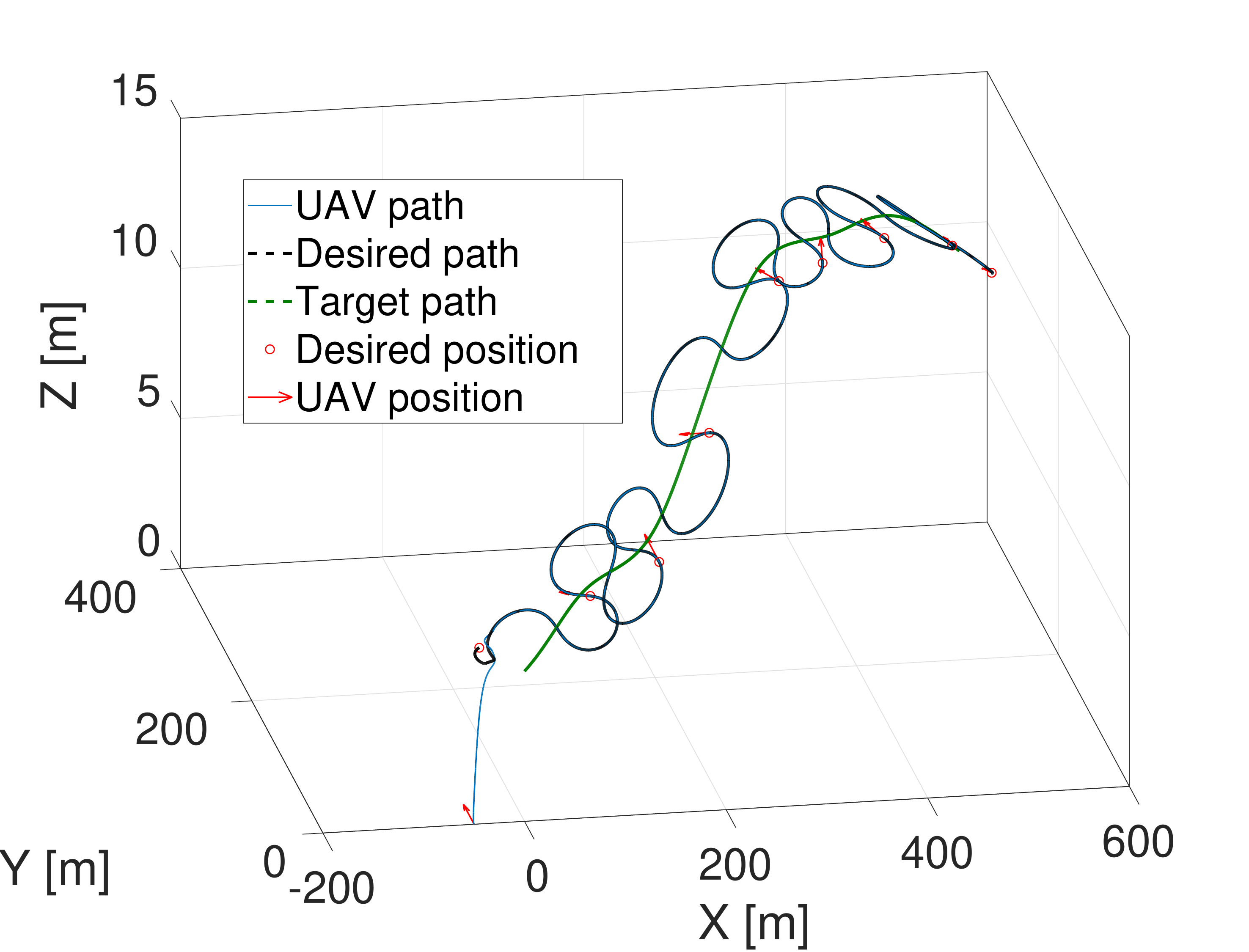}
		\caption{}
		\label{fig:xyz}
	\end{subfigure}%
	\begin{subfigure}[t]{.5\textwidth}
		\centering
		\includegraphics[width=1.0\linewidth]{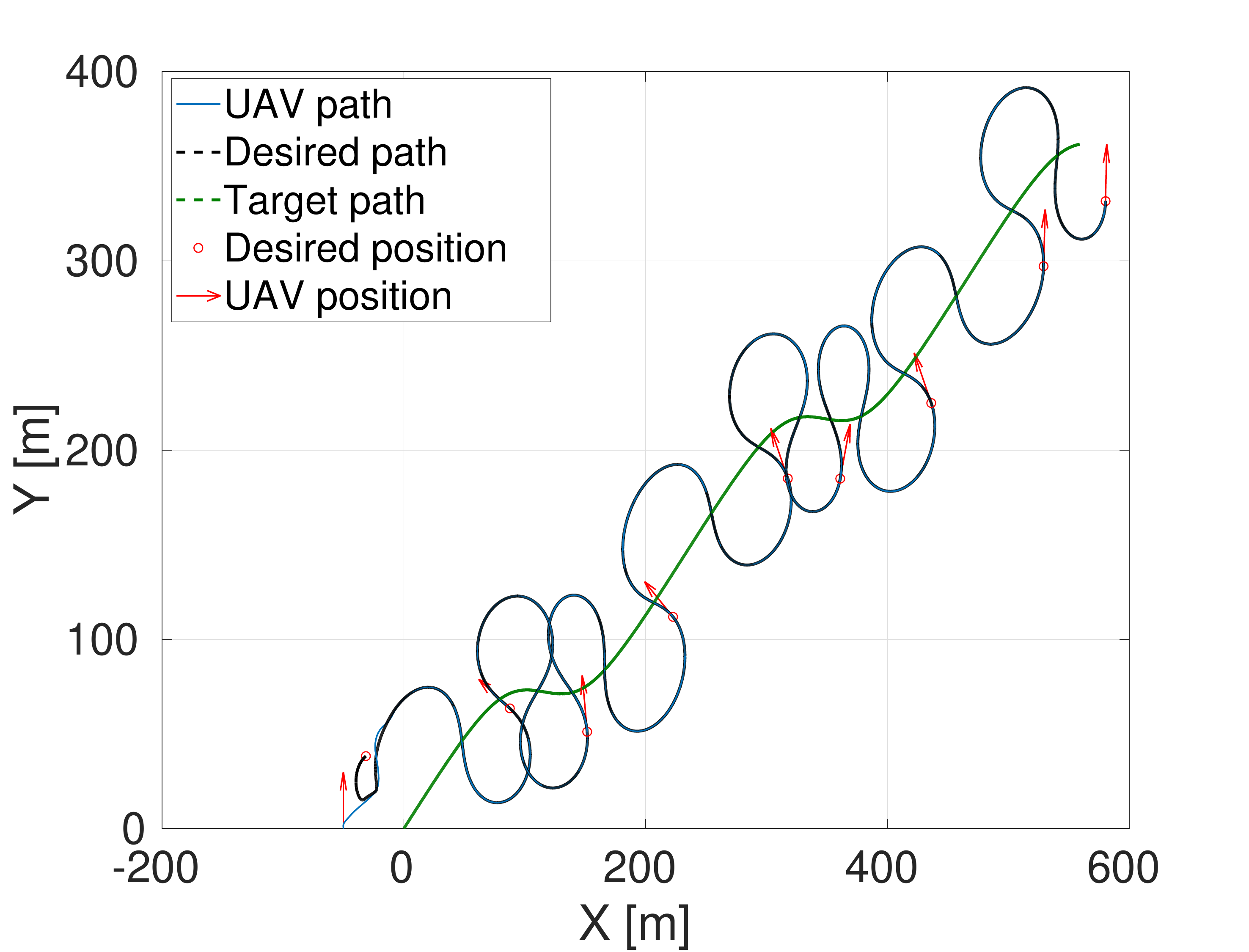}
		\caption{}
		\label{fig:xy}
	\end{subfigure}\\
	\begin{subfigure}[t]{.5\textwidth}
		\centering
		\includegraphics[width=1.0\linewidth]{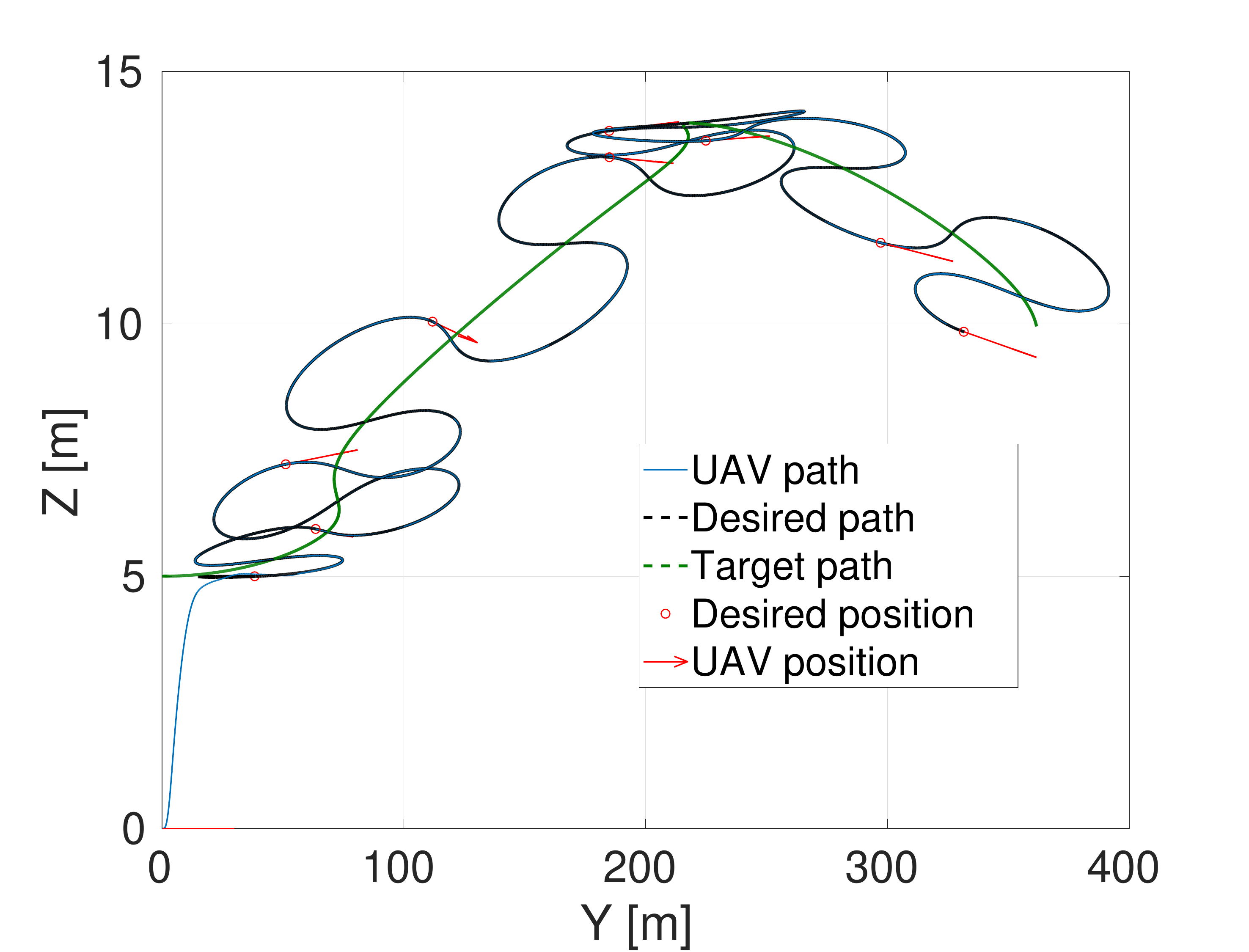}
		\caption{}
		\label{fig:yz}  
	\end{subfigure} %
	\begin{subfigure}[t]{.5\textwidth}
	\centering
	\includegraphics[width=1.0\linewidth]{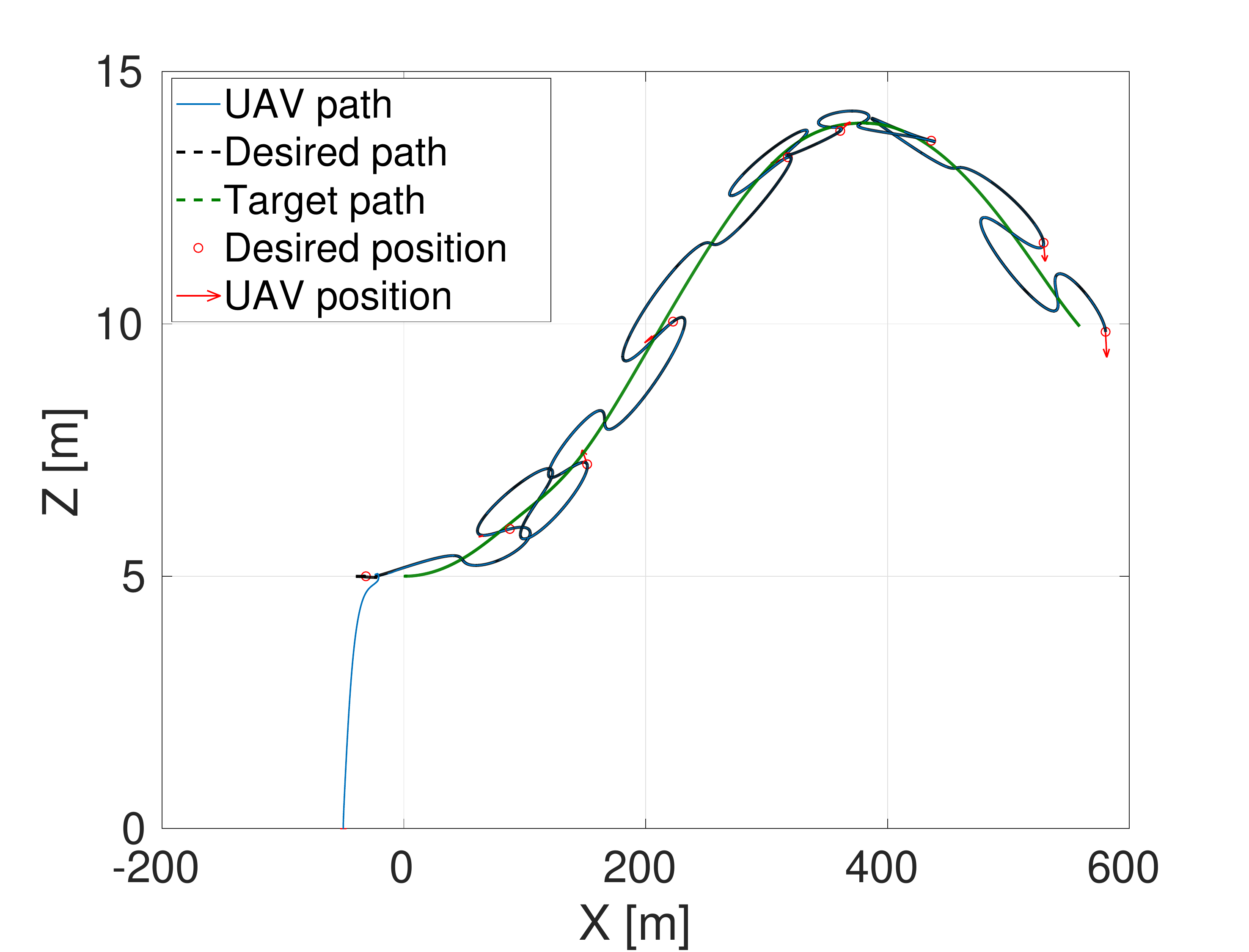}
	\caption{}
	\label{fig:xz}  
\end{subfigure} %
	\label{fig:mpf_position}
	\caption{Position of the target maneuvering in 3D and an UAV following a lemniscate path around the target}
\end{figure*}
\begin{figure*}[h]
	\begin{subfigure}[t]{.3334\textwidth}
		\centering
		\includegraphics[width=1.0\linewidth]{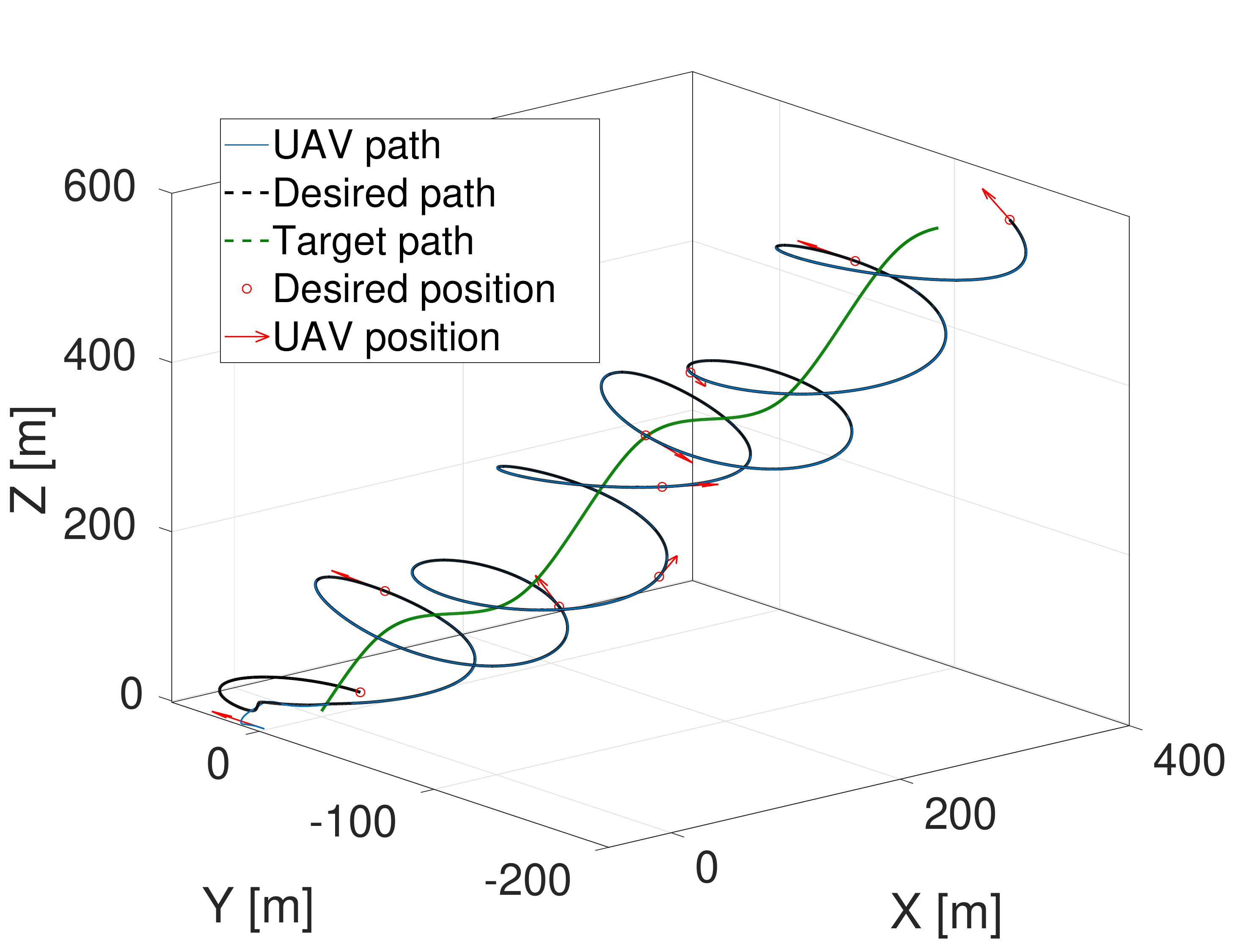}
		\caption{}
		\label{fig:xyz_circle}
	\end{subfigure}%
	\begin{subfigure}[t]{.3334\textwidth}
		\centering
		\includegraphics[width=1.0\linewidth]{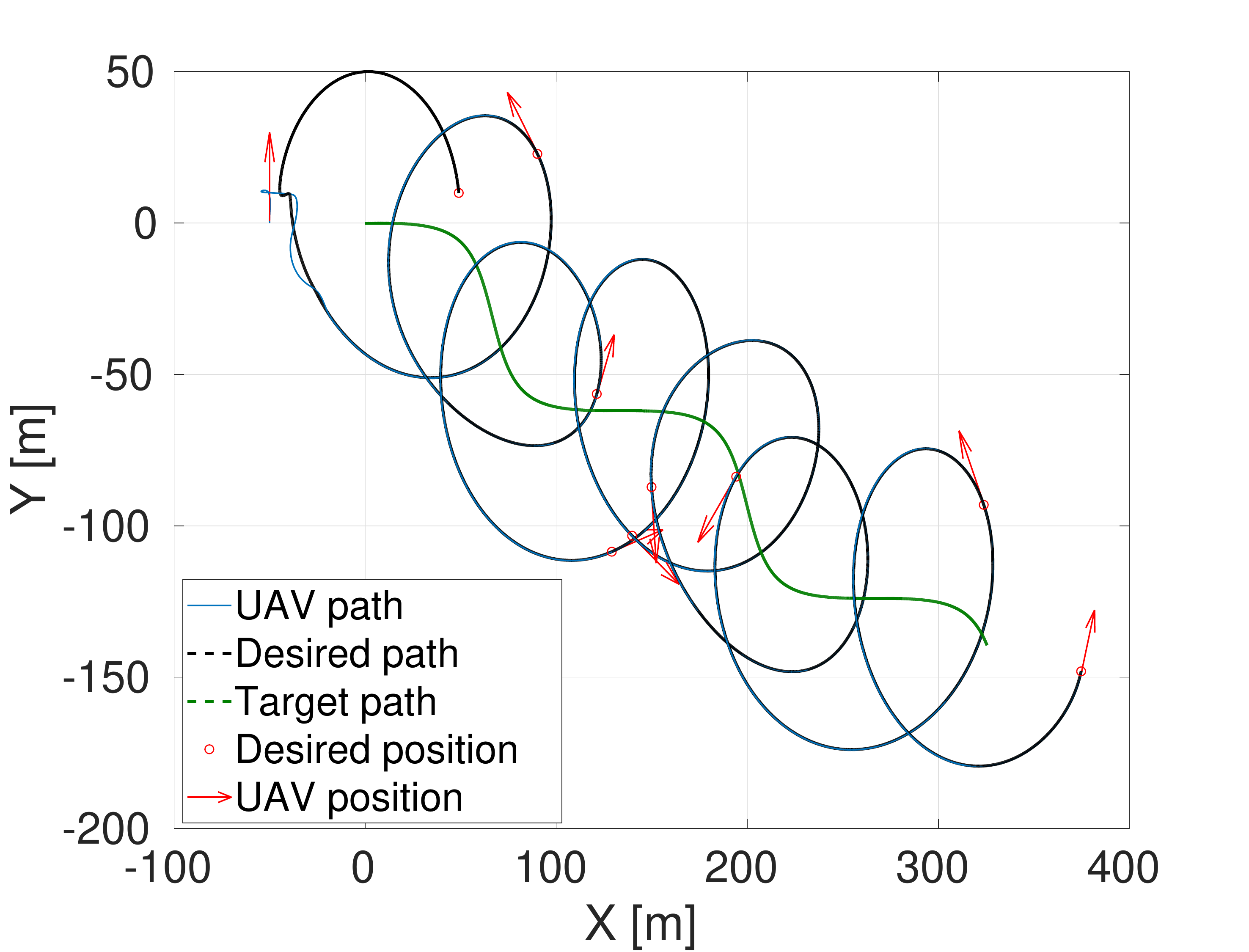}
		\caption{}
		\label{fig:xy_circle}
	\end{subfigure}%
	\begin{subfigure}[t]{.3334\textwidth}
		\centering
		\includegraphics[width=1.0\linewidth]{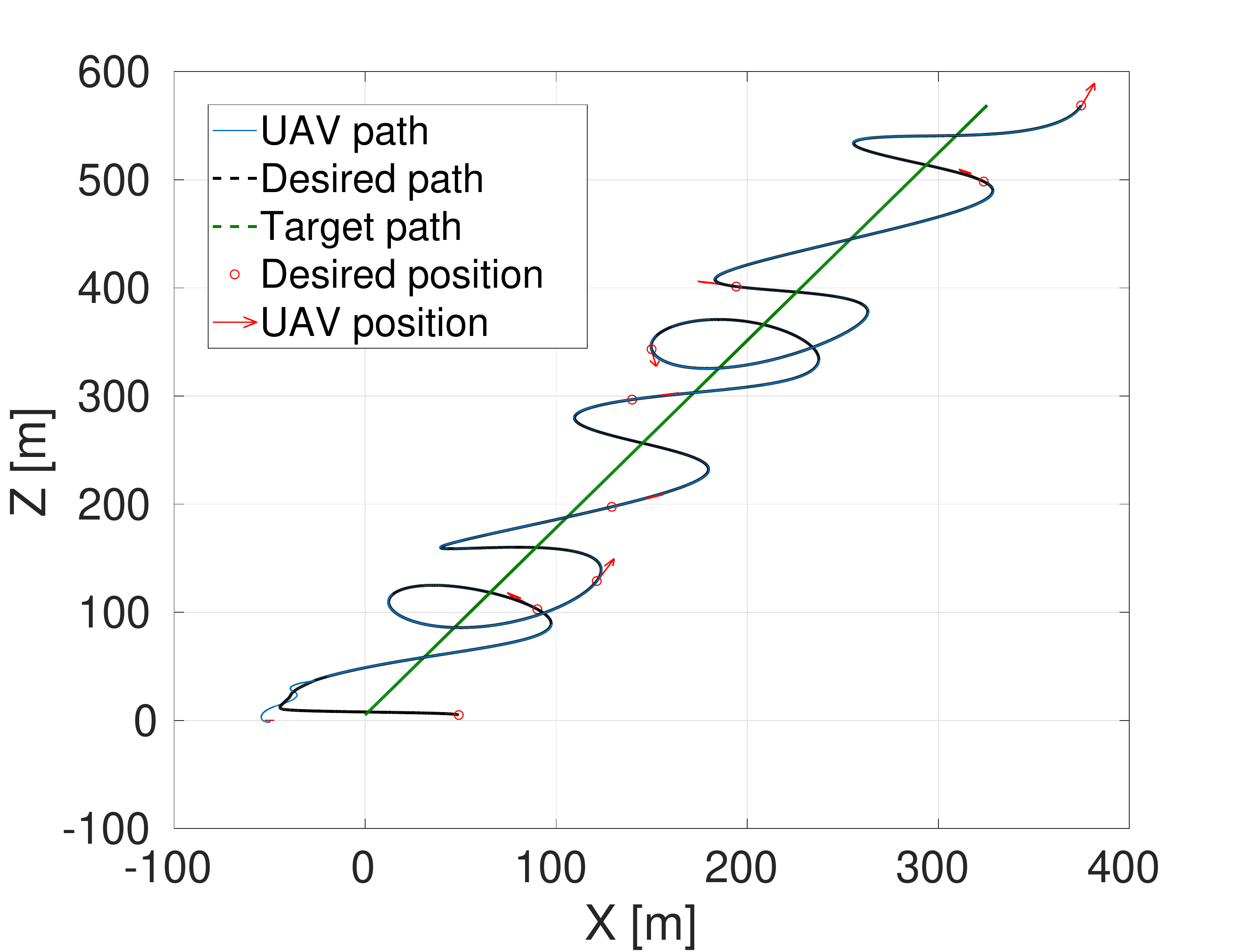}
		\caption{}
		\label{fig:xz_circle}  
	\end{subfigure} %
	\label{fig:mpf_circle}
	\caption{Position of the target maneuvering in 3D and an UAV following a circular path around the target}
\end{figure*}
\begin{figure*}[h]
		\begin{subfigure}[t]{.3334\textwidth}
			\centering
			\includegraphics[width=1.0\linewidth]{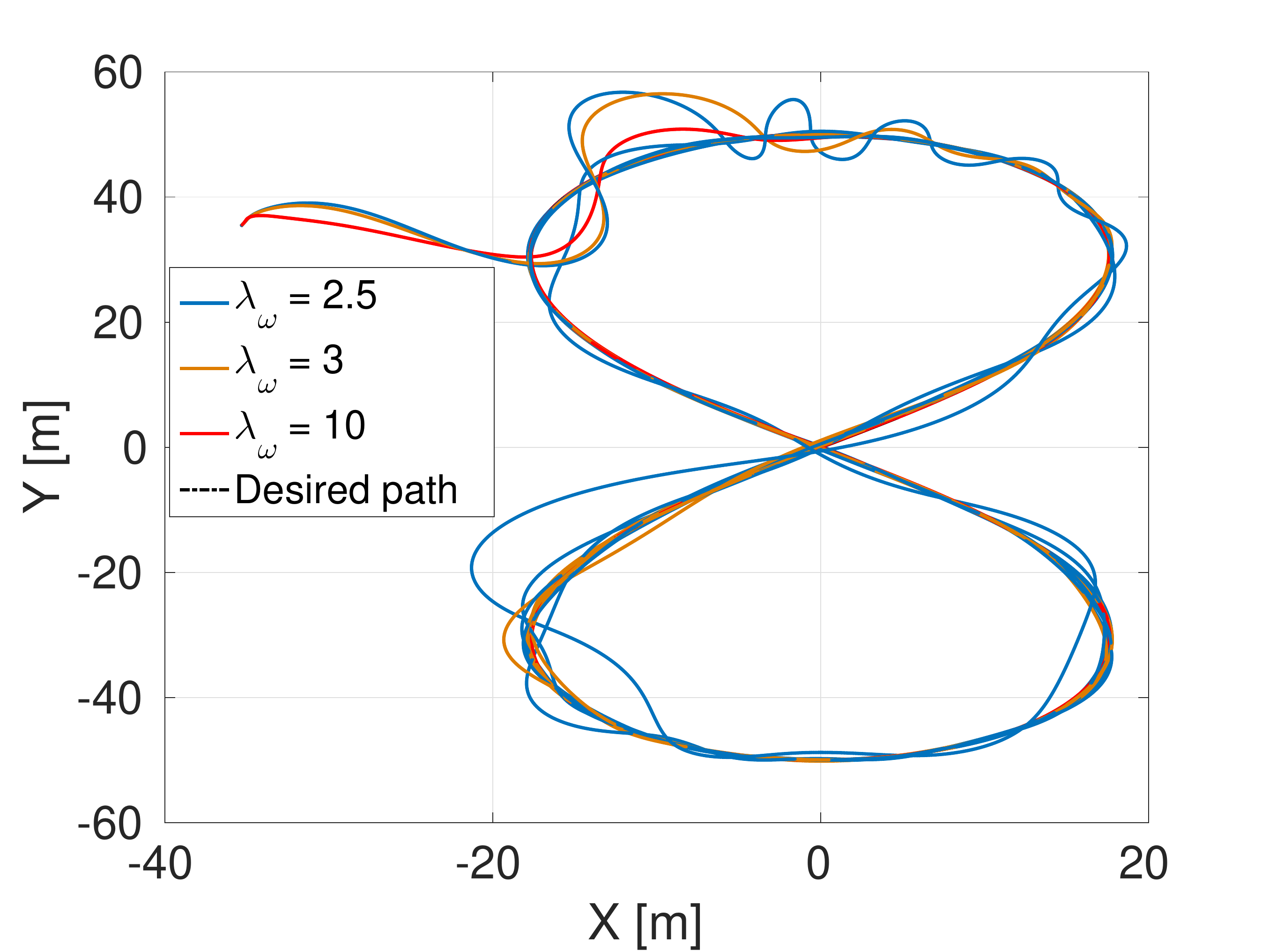}
			\caption{}
			\label{fig:xyz_target}
		\end{subfigure}%
	\begin{subfigure}[t]{.3334\textwidth}
		\centering
		\includegraphics[width=1.0\linewidth]{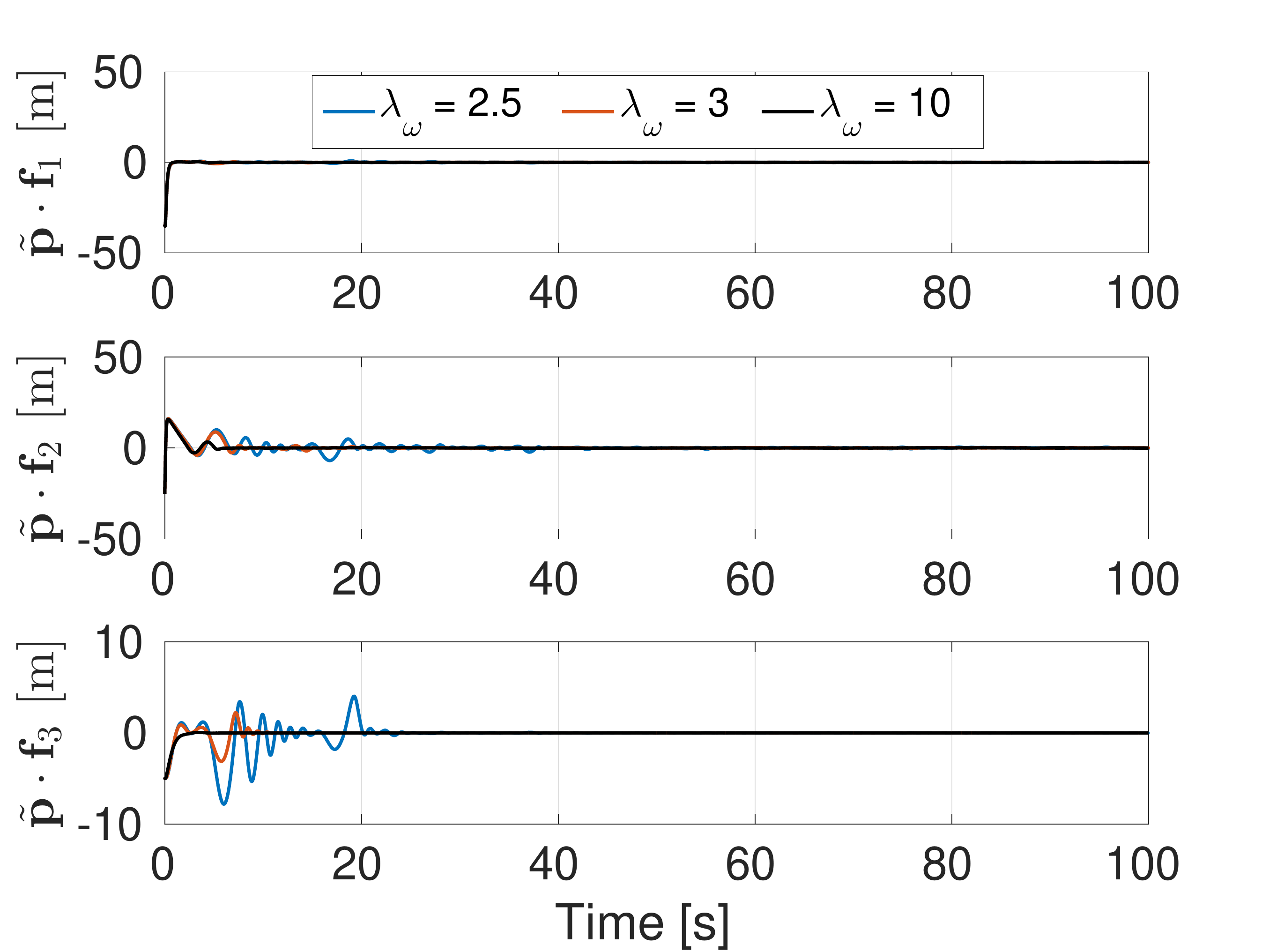}
		\caption{}
		\label{fig:mpferr}
	\end{subfigure}%
	\begin{subfigure}[t]{.3334\textwidth}
		\centering
		\includegraphics[width=1.0\linewidth]{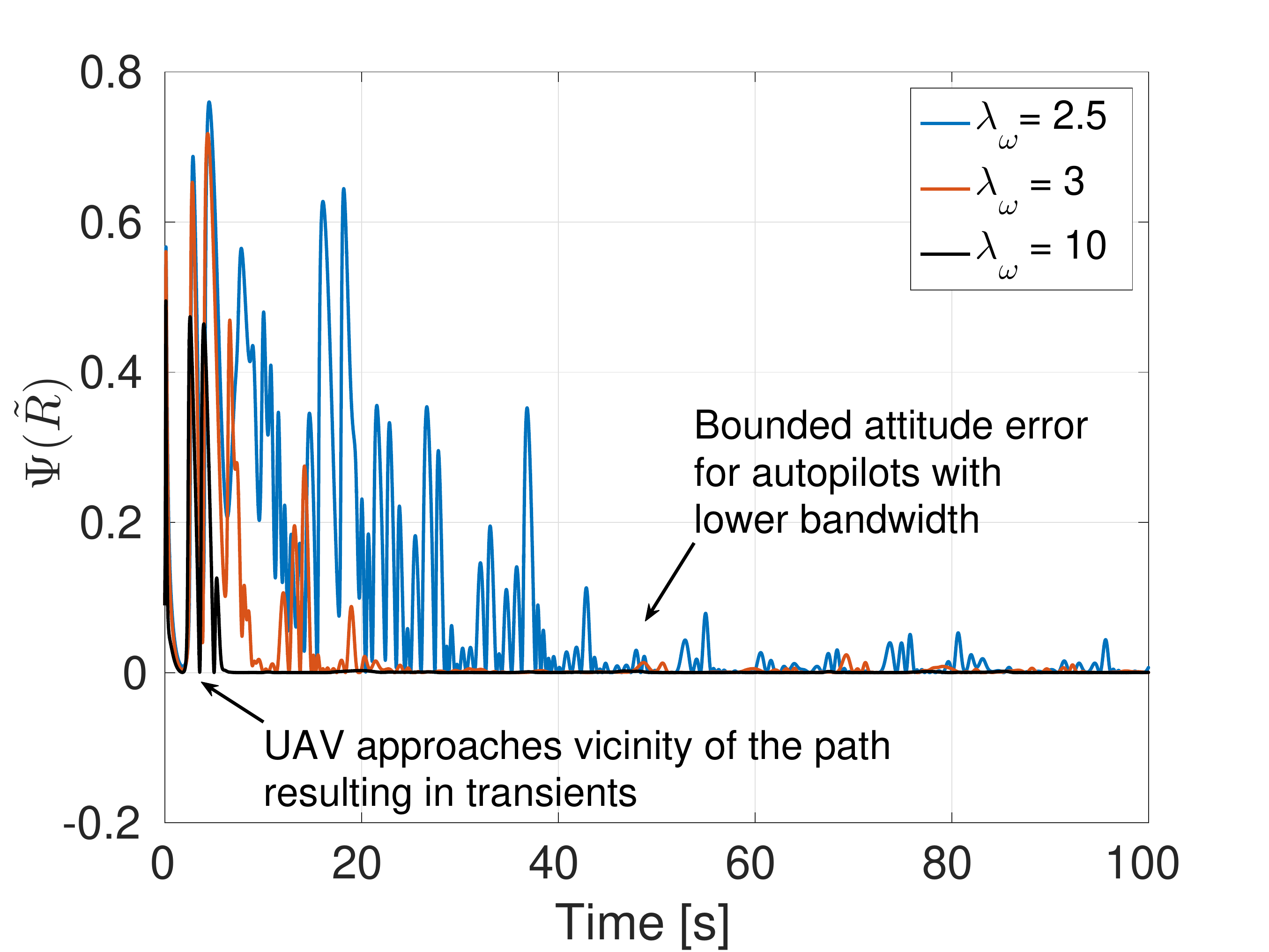}
		\caption{}
		\label{fig:atterr}  
	\end{subfigure}\\
	\begin{subfigure}[t]{.3334\textwidth}
	\centering
	\includegraphics[width=1.0\linewidth]{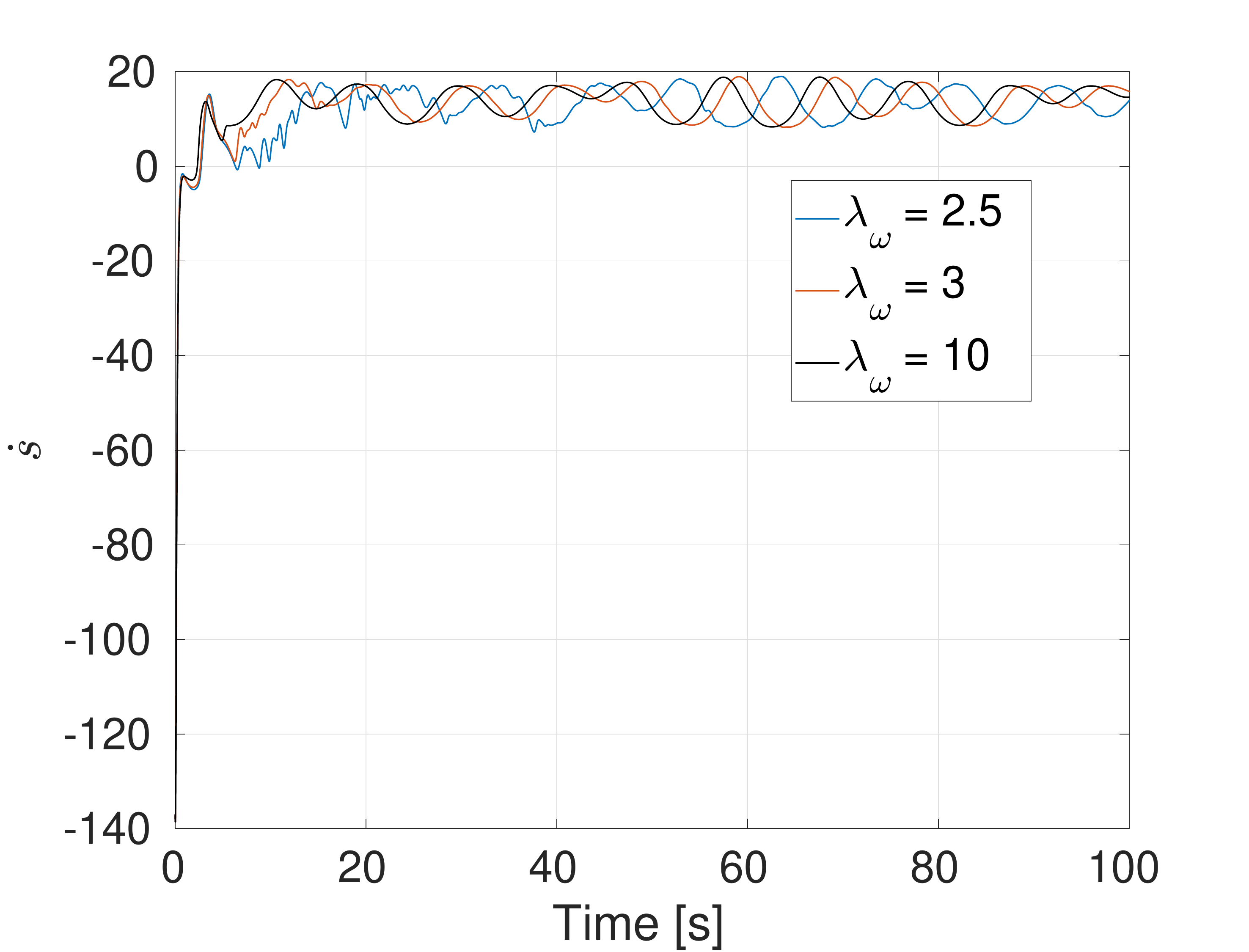}
	\caption{}
	\label{fig:sDot}
\end{subfigure}%
\begin{subfigure}[t]{.3334\textwidth}
	\centering
	\includegraphics[width=1.0\linewidth]{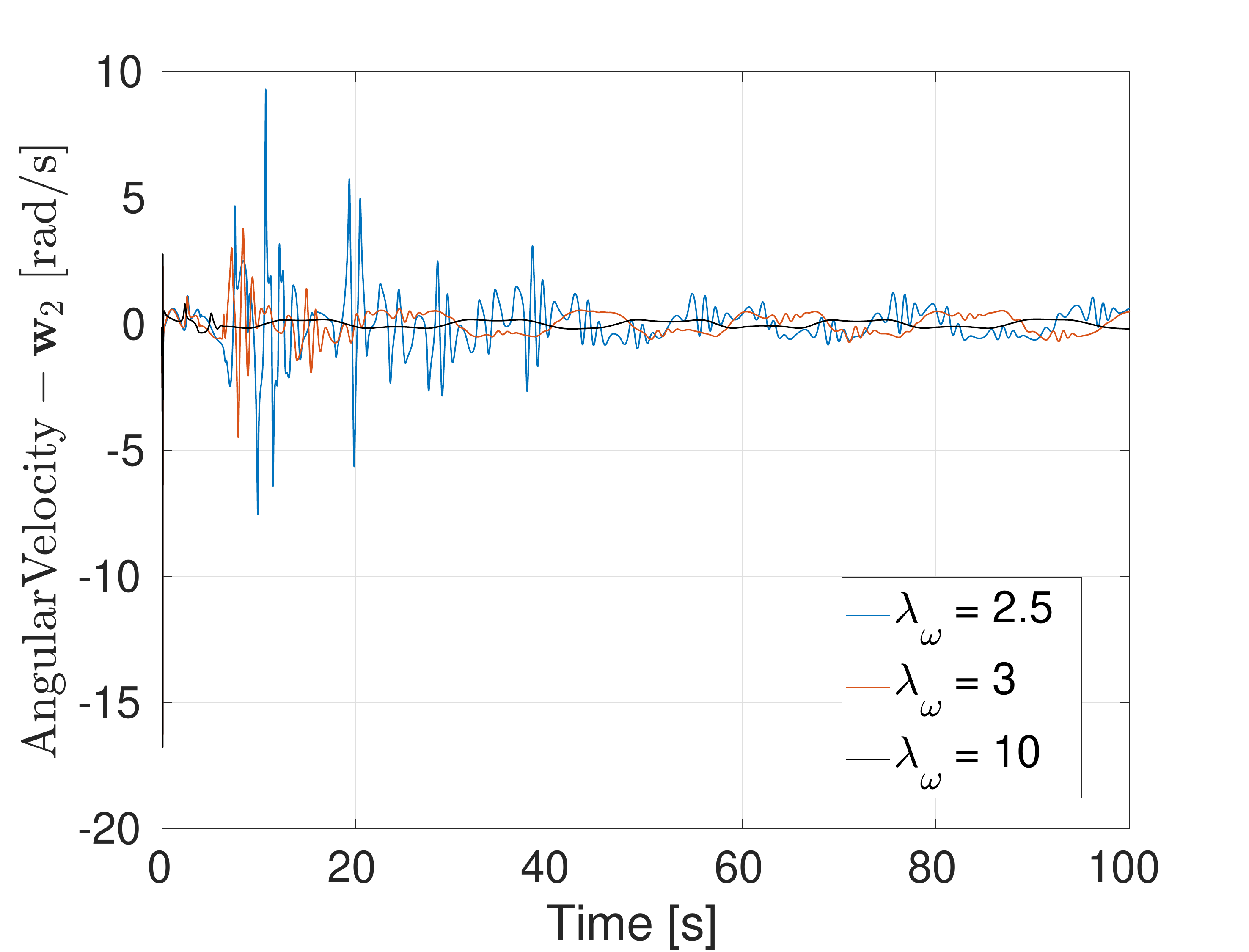}
	\caption{}
	\label{fig:w2}
\end{subfigure}%
\begin{subfigure}[t]{.3334\textwidth}
	\centering
	\includegraphics[width=1.0\linewidth]{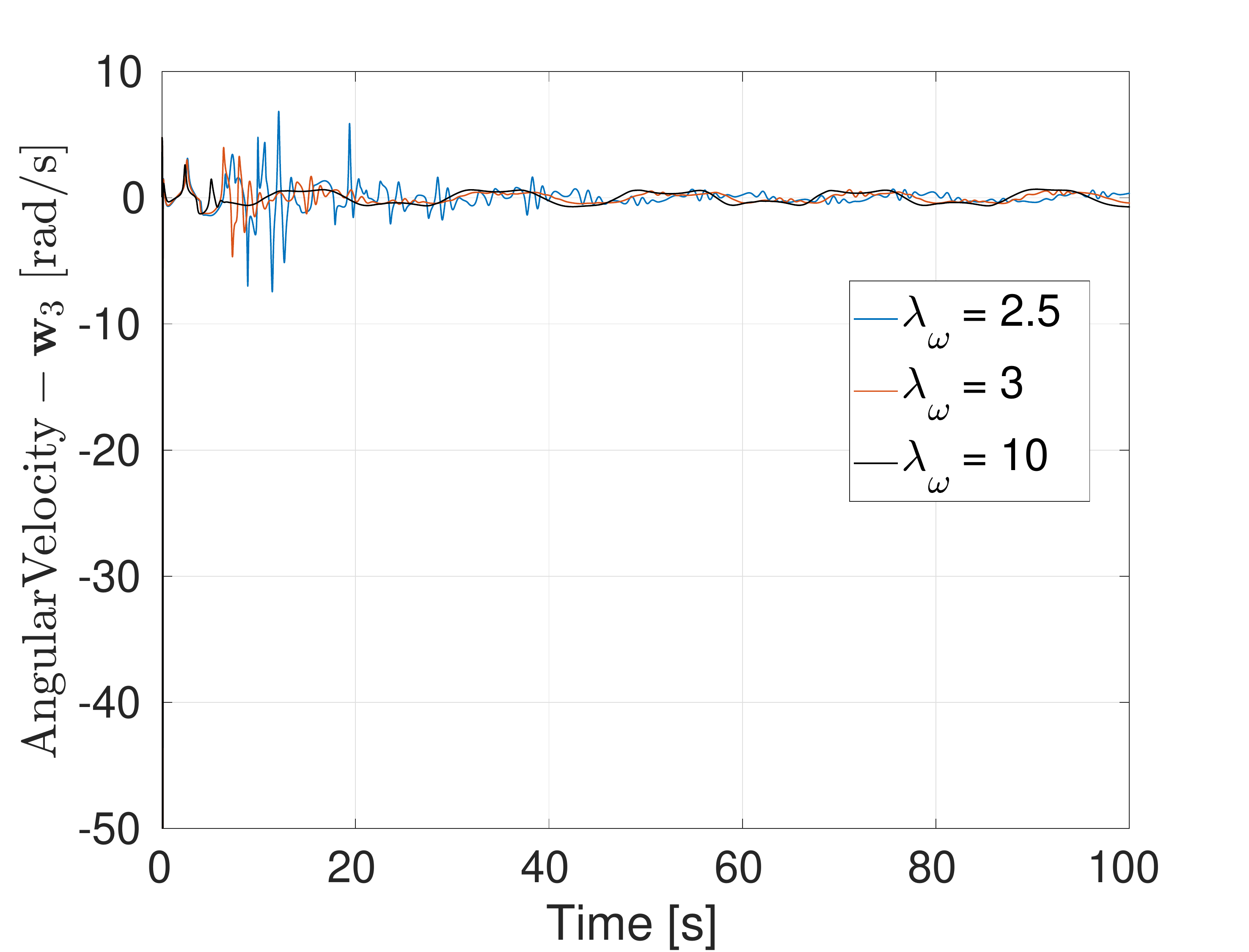}
	\caption{}
	\label{fig:w3}  
\end{subfigure} %
	\label{fig:mpf}
	\caption{Plot of (a) Target relative desired path and UAV path, (b) MPF error $\tilde{\mathbf{p}}$, and (c) Attitude error function $\Psi(\tilde{R})$ (d) Virtual control input $\dot{s}$, (e) Angular velocity reference about $\mathbf{w}_2$, and (f) Angular velocity reference about $\mathbf{w}_3$}
\end{figure*}
\begin{figure*}[h]
	\begin{subfigure}[t]{.5\textwidth}
		\centering
		\includegraphics[width=1.0\linewidth]{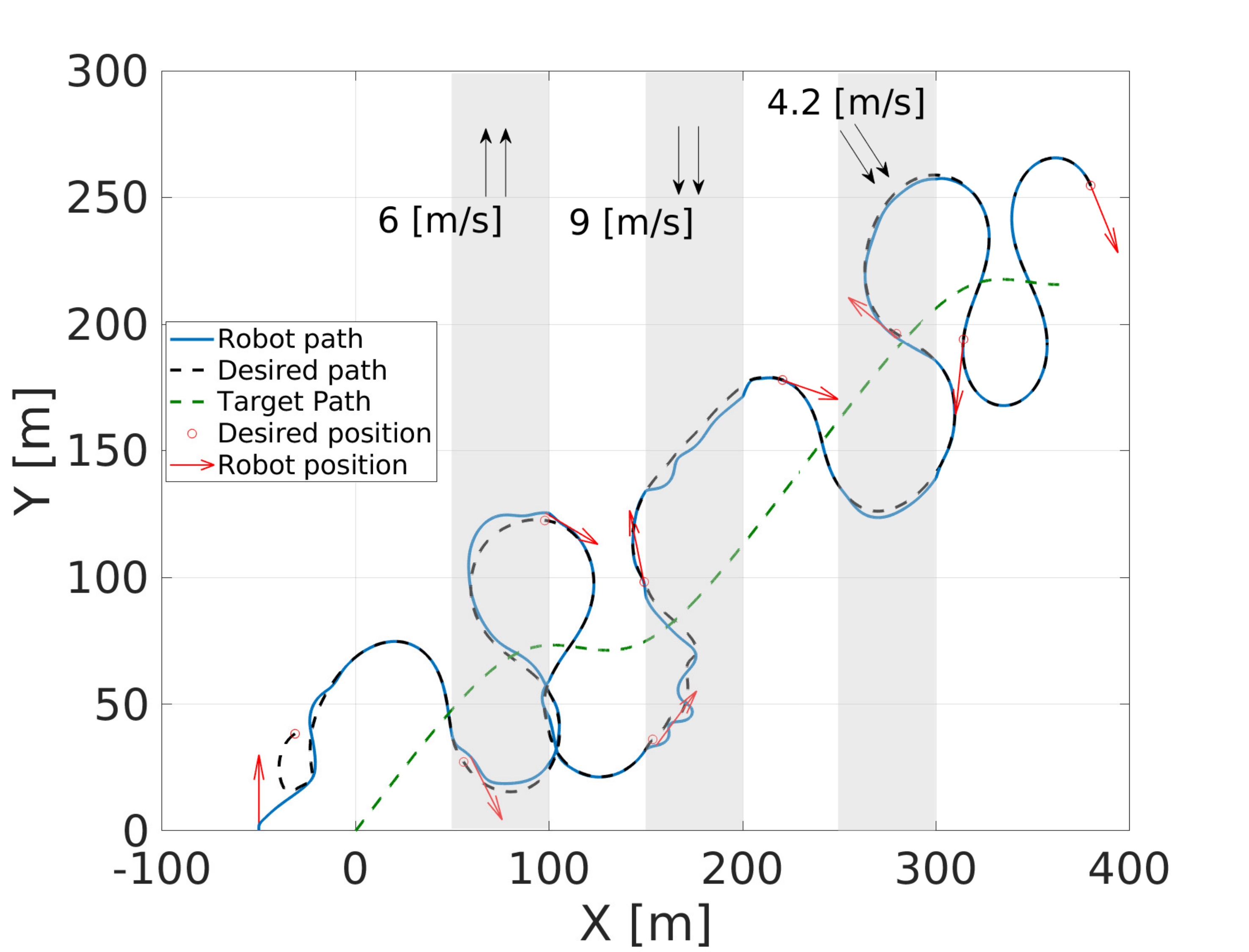}
		\caption{MPF simulation with varied wind disturbances}
		\label{fig:xy_wind_spatial}
	\end{subfigure}%
	\begin{subfigure}[t]{.5\textwidth}
		\centering
		\includegraphics[width=1.0\linewidth]{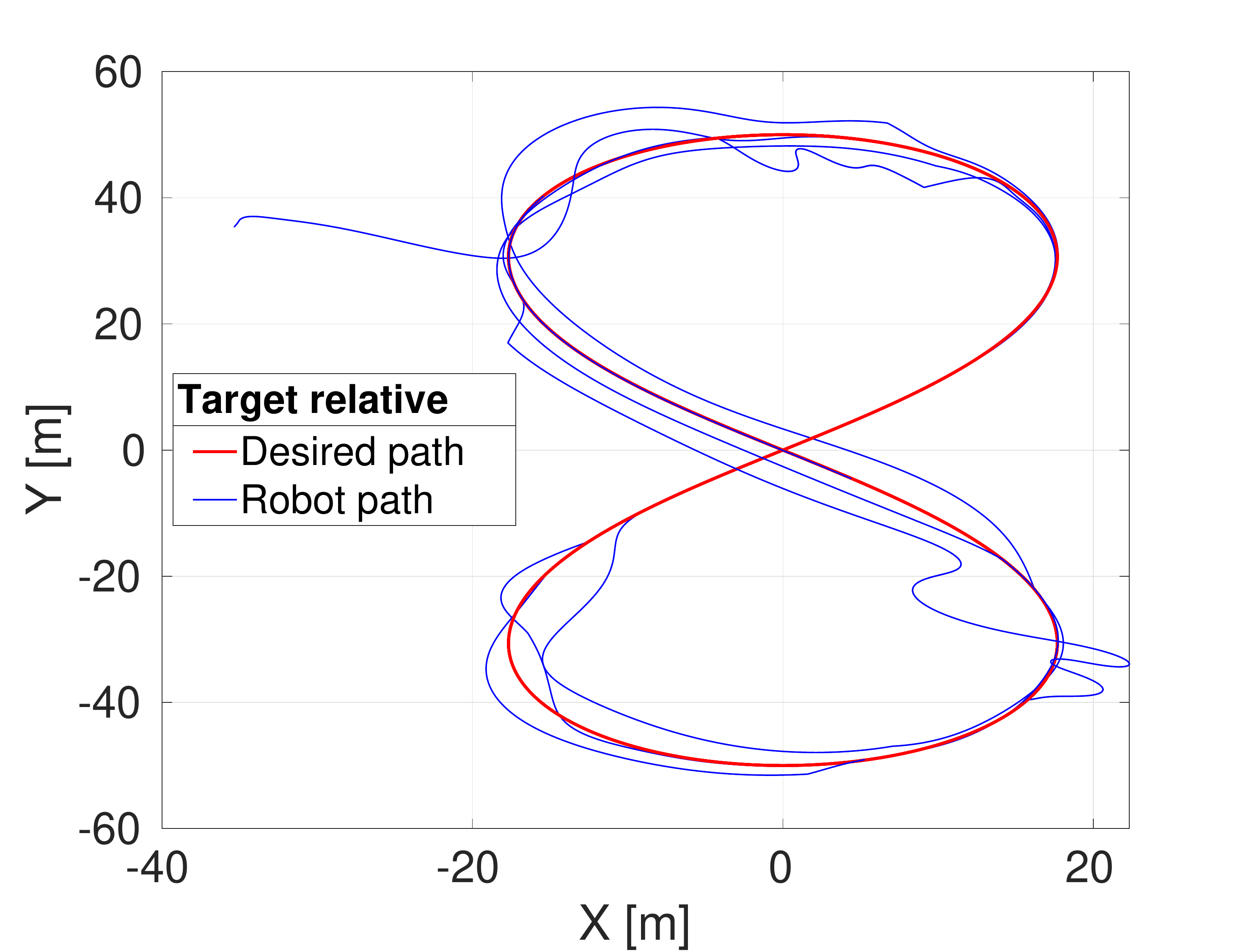}
		\caption{Target relative robot position}
		\label{fig:target_relative_wind_spatial}
	\end{subfigure}\\
	\begin{subfigure}[t]{.5\textwidth}
	\centering
	\includegraphics[width=1.0\linewidth]{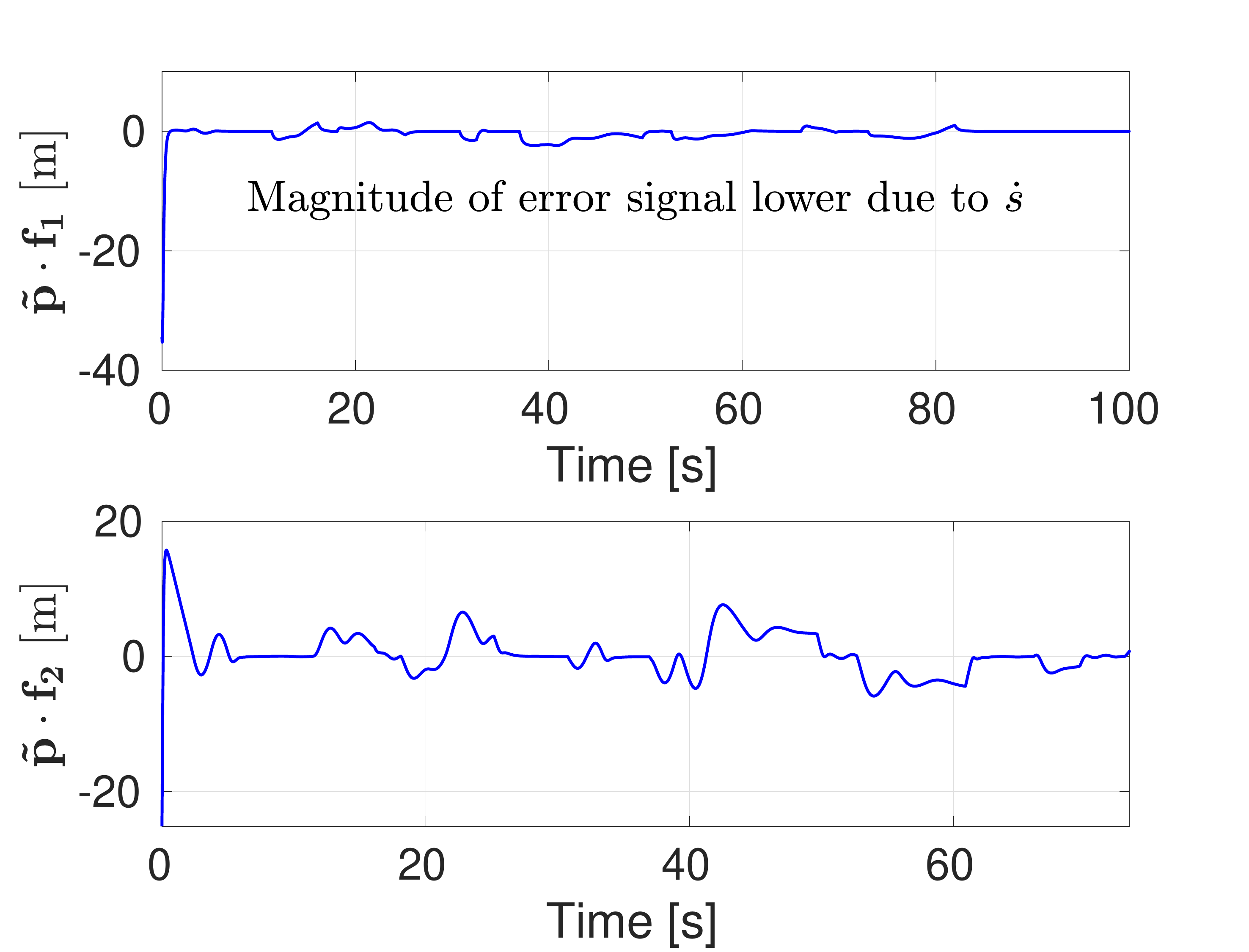}
	\caption{MPF error}
	\label{fig:pos_error_wind_spatial}
\end{subfigure}%
\begin{subfigure}[t]{.5\textwidth}
	\centering
	\includegraphics[width=1.0\linewidth]{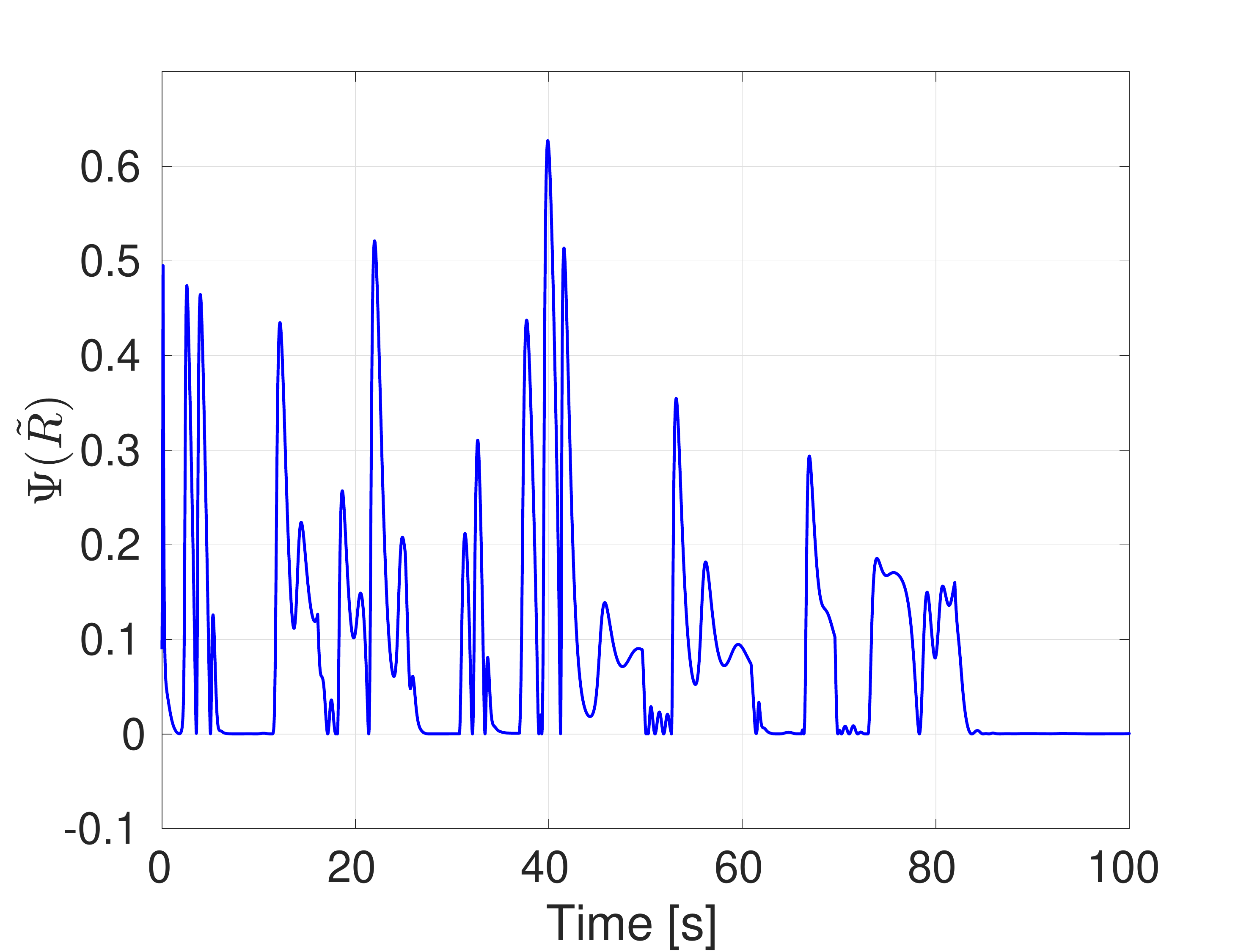}
	\caption{Attitude error function}
	\label{fig:att_err_wind_spatial}
\end{subfigure}
	\label{fig:mpf_wind}
	\caption{Position of the target maneuvering in 3D and an UAV following a lemniscate path around the target in the presence of wind disturbances.}
\end{figure*}
\textcolor{black}{\begin{remark}
		Consider an external disturbance signal $\boldsymbol{\eta}$ acting on the robot position as $\dot{\mathbf{p}} = v_w\mathbf{w}_1 + \boldsymbol{\eta}$. Further, assume that there exists a disturbance estimator that provides an estimate of the disturbance signal $\hat{\boldsymbol{\eta}} = \boldsymbol{\eta} - \tilde{\boldsymbol{\eta}}$, where $\tilde{\boldsymbol{\eta}}$ is a bounded  disturbance estimation error signal. Then, introducing the term $\hat{\boldsymbol{\eta}}$ within the parenthesis of Equation \eqref{eq:sdot} and in the computation of Equation \eqref{eq:property_equation} and consequently in Equation \eqref{eq:wdbasis}, it is possible to compensate for the external disturbances and demonstrate the ISS of the proposed MPF law with respect to the disturbance estimation error signal $\tilde{\boldsymbol{\eta}}$. The proof holds with straightforward manipulation of expressions within the proof. It is thus, possible to conclude that proposed MPF control law is robust to the bounded disturbance signals (e.g., wind), and in particular that the MPF error signals remain bounded and will converge to a residual error that depends on the size of the steady state disturbance estimation error signal $\tilde{\boldsymbol{\eta}}$. Notice also that exclusion of the compensation term for the external disturbances in the control laws implies  $\hat{\boldsymbol{\eta}} = 0$, which means that in this case $\tilde{\boldsymbol{\eta}} = \boldsymbol{\eta}$. This is indeed what we do in the simulation results described in the next section.
\end{remark}}
\section{Simulation Results}\label{sec:simulationresults}
The proposed MPF control strategy was validated in simulations for a generic scenario where an UAV is tracking a target that is executing a 3D maneuver with linear and angular velocities given by $\dot{\mathbf{p}}_t^T = [3\sqrt{(\cos(0.1t) + 1)^2 + 1}, \hspace{0.1cm} 0, \hspace{0.1cm} 0]^\prime$ and $\boldsymbol{\omega}_{TI}^T = \left[0, \hspace{0.1cm} -0.001\cos(0.03t), \hspace{0.1cm} \frac{-\sin(0.1t)}{10[(\cos(0.1t) + 1)^2 + 1]^{3/2}}\right]^\prime$. A lemniscate path is chosen as desired path fixed with respect to the target frame given by $\mathbf{p}_d^T(s) = \left[\frac{50\sin(0.01s)\cos(0.01s)}{\sin^2(0.01s) + 1}, \hspace{0.1cm} \frac{50\cos(0.01s)}{\sin^2(0.01s) + 1}, \hspace{0.1cm} 0\right]^\prime$. The desired speed of the UAV is chosen as 15 [m/s], with controller gains $K_p = 4$, $K_{\tilde{R}} = 2$ and $\alpha = 1$. The initial position of the UAV is set to $\mathbf{p} = [-50, \hspace{0.1cm} 0, \hspace{0.1cm} 0]^\prime$ and initial attitude of $R_W^I = \mathrm{Rot}(0,0,\frac{\pi}{2})$, where the arguments of the function $\mathrm{Rot}()$ specifies rotation about the axes of vehicle frame $\{W\}$. Similarly, the target initial position was set to $\mathbf{p}_t = [0, \hspace{0.1cm} 0, \hspace{0.1cm} 5]^\prime$ with initial attitude $R_T^I = \mathrm{Rot}(0,0,\frac{\pi}{4})$. \textcolor{black}{The autopilot controller was simulated as a first order system i.e., $\dot{\boldsymbol{\omega}} = -\lambda_\omega\boldsymbol{\omega} + \lambda_\omega\bar{\boldsymbol{\omega}}$, where $\lambda_\omega > 0$ relates to the bandwidth of the autopilot, $\boldsymbol{\omega} = \Pi_R\boldsymbol{\omega}_{WI}^W$, and $\bar{\boldsymbol{\omega}} = \Pi_R\bar{\boldsymbol{\omega}}_{WI}^W$ with slight abuse of notation. The performance of MPF controller was simulated for the values of $\lambda_\omega = \{2.5, 3, 10\}$.}

\textcolor{black}{Figure \ref{fig:xyz} illustrates the plot of UAV position and the target position for the proposed scenario for $\lambda_\omega = 10$. Different planar perspectives are provided in Figure \ref{fig:xy}, \ref{fig:yz} and \ref{fig:xz}. In order to illustrate the generic nature of the controller, a simulation with circular path is also presented in Figure \ref{fig:xyz_circle}, \ref{fig:xy_circle} and \ref{fig:xz_circle}. From the simulations it can be noted that the MPF control law is able to generate reference commands for the autopilot controller that allows the robot to achieve the MPF objective. Reference \cite{jain2018auv} also present the case study where in the target motion is estimated. A simulation video can also be viewed at https://youtu.be/G-JKqtytyFM.}

\textcolor{black}{Figure \ref{fig:xyz_target} shows the plot of position of the robotic vehicle relative to the target with autopilot system of different bandwidth. The MPF position error converges to the vicinity of zero as can be seen in Figure \ref{fig:mpferr}. The degradation in the performance of the controller is evident for lower bandwidth autopilot. However, the MPF position error remains bounded. Similarly, the MPF attitude error converges to neighborhood of zero and remains bounded as can be seen from Figure \ref{fig:atterr} that shows the plot of the value of attitude error function $\Psi(\tilde{R})$. Figure \ref{fig:sDot} denotes the plot of the virtual control input that enables faster convergence of the robot to the virtual point that moves along the path. Angular velocity references generated about $\mathbf{w}_2$ and $\mathbf{w}_3$ axis are presented in Figure \ref{fig:w2} and \ref{fig:w3} respectively. As mentioned, the autopilots track these signals according with a first order system with bandwidth  $\lambda_\omega$. Note that the angular velocity along the $\mathbf{w}_1$ axis is always zero by the nature of the control design. The roll commands are generated by the autopilots depending on the type of the robotic vehicle. Although, we illustrate the results with an UAV, it is important to stress that the approach is generic and valid for any robotic system with an existing autopilot. The feasibility of the target motion and the specified path can be viewed as trajectory or path planning problem that is not addressed in this work. Here we assume that the specifications of the MPF system are provided such that they are feasible for the robotic mission. The simulations validate the ISS of the proposed MPF controller and it should be noted that the position and attitude errors converge to zero with high bandwidth autopilot controllers. Lower values of $\lambda_\omega$ can be used for targets that maneuver slower than the scenario considered in this paper.}

\textcolor{black}{The performance of the proposed control law is further tested by simulating the system with disturbance such as wind gusts. For simplicity the wind gusts are considered for the planar case. Figure \ref{fig:xy_wind_spatial} shows the plot for wind gusts acting at specified regions. The wind gusts are of considerable magnitude, given the air speed was set to $v_w = 15$ [m/s]. It can be noted that the robot follows the desired moving path successfully, albeit with non zero MPF position errors. However the errors are limited to the neighborhood of the origin as can be seen from Figure \ref{fig:pos_error_wind_spatial}. This validates the ISS property of the proposed control law as can be seen from Figure \ref{fig:target_relative_wind_spatial}. The video for simulations with wind gusts can be viewed at https://youtu.be/WcskiZlCzZ0. The performance can certainly be improved by incorporating the disturbance estimators as done in \cite{aguiar2007trajectory, guan2019mpf,zheng2020mpf}. Additionally, note that the estimate of the region of attraction, conditions on the gains and the performance bounds on the autopilot are conservative. Consequently, it could be verified in simulations that the MPF position and attitude errors converge to the neighborhood of zero for various other initial conditions.} 
 
\section{Conclusion}\label{sec:conclusion}
This paper presented a 3D, geometric MPF control law on SO(3) for the robotic vehicles such as fixed-wing UAVs/AUVs that require a minimum forward speed for its operation. The convergence to the moving path was achieved by controlling the angular velocity of the vehicle, aided by the design of a virtual control input that controls the progression of the virtual point moving along the path. The proposed control strategy was proven to be Input-to-State Stable within an estimate of a region of attraction and its efficacy was illustrated in simulations in the presence of imperfect tracking of the MPF control commands by the autopilot inner-loop controller. The efficacy of the approach was also supported by simulations in the presence of the constant wind disturbances. Future work involves, experimental validation of the presented control law in presence of the disturbances and estimation of the relevant target information. To this end, robustness aspects could be considered through the design of disturbance estimators.





\bibliographystyle{IEEEtran}
\bibliography{IEEEabrv,mybib}

\begin{thebibliography}{10}
\providecommand{\url}[1]{#1}
\csname url@samestyle\endcsname
\providecommand{\newblock}{\relax}
\providecommand{\bibinfo}[2]{#2}
\providecommand{\BIBentrySTDinterwordspacing}{\spaceskip=0pt\relax}
\providecommand{\BIBentryALTinterwordstretchfactor}{4}
\providecommand{\BIBentryALTinterwordspacing}{\spaceskip=\fontdimen2\font plus
\BIBentryALTinterwordstretchfactor\fontdimen3\font minus
  \fontdimen4\font\relax}
\providecommand{\BIBforeignlanguage}[2]{{%
\expandafter\ifx\csname l@#1\endcsname\relax
\typeout{** WARNING: IEEEtran.bst: No hyphenation pattern has been}%
\typeout{** loaded for the language `#1'. Using the pattern for}%
\typeout{** the default language instead.}%
\else
\language=\csname l@#1\endcsname
\fi
#2}}
\providecommand{\BIBdecl}{\relax}
\BIBdecl

\bibitem{micaelli1993trajectory}
A.~Micaelli and C.~Samson, ``Trajectory tracking for unicycle-type and
  two-steering-wheels mobile robots,'' Ph.D. dissertation, INRIA, 1993.

\bibitem{samson1993time}
C.~Samson, ``Time-varying feedback stabilization of car-like wheeled mobile
  robots,'' \emph{The International journal of robotics research}, vol.~12,
  no.~1, pp. 55--64, 1993.

\bibitem{samson1992path}
------, ``Path following and time-varying feedback stabilization of a wheeled
  mobile robot,'' in \emph{Proceedings of the international conference on
  advanced robotics and computer vision}, vol.~13, 1992, pp. 1--1.

\bibitem{aguiar2005path}
A.~P. Aguiar, J.~P. Hespanha, and P.~V. Kokotovic, ``Path-following for
  nonminimum phase systems removes performance limitations,'' \emph{IEEE
  Transactions on Automatic Control}, vol.~50, no.~2, pp. 234--239, 2005.

\bibitem{aguiar2007trajectory}
A.~P. Aguiar and J.~P. Hespanha, ``Trajectory-tracking and path-following of
  underactuated autonomous vehicles with parametric modeling uncertainty,''
  \emph{IEEE Transactions on Automatic Control}, vol.~52, no.~8, pp.
  1362--1379, Aug 2007.

\bibitem{ENCARNACAO2000507}
\BIBentryALTinterwordspacing
P.~Encarnação, A.~Pascoal, and M.~Arcak, ``Path following for marine vehicles
  in the presence of unknown currents1,'' \emph{IFAC Proceedings Volumes},
  vol.~33, no.~27, pp. 507 -- 512, 2000, 6th IFAC Symposium on Robot Control
  (SYROCO 2000), Vienna, Austria, 21-23 September 2000. [Online]. Available:
  \url{http://www.sciencedirect.com/science/article/pii/S1474667017379806}
\BIBentrySTDinterwordspacing

\bibitem{BELLETER2016588}
\BIBentryALTinterwordspacing
D.~Belleter, C.~Paliotta, M.~Maggiore, and K.~Pettersen, ``Path following for
  underactuated marine vessels,'' \emph{IFAC-PapersOnLine}, vol.~49, no.~18,
  pp. 588 -- 593, 2016, 10th IFAC Symposium on Nonlinear Control Systems NOLCOS
  2016. [Online]. Available:
  \url{http://www.sciencedirect.com/science/article/pii/S2405896316318092}
\BIBentrySTDinterwordspacing

\bibitem{xargay2013time}
E.~Xargay, I.~Kaminer, A.~Pascoal, N.~Hovakimyan, V.~Dobrokhodov, V.~Cichella,
  A.~Aguiar, and R.~Ghabcheloo, ``Time-critical cooperative path following of
  multiple unmanned aerial vehicles over time-varying networks,'' \emph{Journal
  of Guidance, Control, and Dynamics}, vol.~36, no.~2, pp. 499--516, 2013.

\bibitem{kaminer2017time}
I.~Kaminer, A.~M. Pascoal, E.~Xargay, N.~Hovakimyan, V.~Cichella, and
  V.~Dobrokhodov, \emph{Time-Critical Cooperative Control of Autonomous Air
  Vehicles}.\hskip 1em plus 0.5em minus 0.4em\relax Butterworth-Heinemann,
  2017.

\bibitem{tiago2013ground}
T.~Oliveira and P.~Encarna{\c{c}}{\~a}o, ``Ground target tracking control
  system for unmanned aerial vehicles,'' \emph{Journal of Intelligent \&
  Robotic Systems}, vol.~69, no. 1-4, pp. 373--387, 2013.

\bibitem{tiago2016MPF}
T.~Oliveira, A.~P. Aguiar, and P.~Encarnação, ``Moving path following for
  unmanned aerial vehicles with applications to single and multiple target
  tracking problems,'' \emph{IEEE Transactions on Robotics}, vol.~32, no.~5,
  pp. 1062--1078, Oct 2016.

\bibitem{reis2019robust}
M.~F. Reis, R.~P. Jain, A.~P. Aguiar, and J.~B. de~Sousa, ``Robust moving path
  following control for robotic vehicles: Theory and experiments,'' \emph{IEEE
  Robotics and Automation Letters}, vol.~4, no.~4, pp. 3192--3199, 2019.

\bibitem{jain2018moving}
R.~P. Jain, A.~P. Aguiar, A.~Alessandretti, and J.~Borges~de Sousa, ``Moving
  path following control of constrained underactuated vehicles: A nonlinear
  model predictive control approach,'' in \emph{2018 AIAA Information
  Systems-AIAA Infotech@ Aerospace}, 2018, p. 0509.

\bibitem{khalil}
\BIBentryALTinterwordspacing
H.~K. Khalil, \emph{Nonlinear systems}.\hskip 1em plus 0.5em minus 0.4em\relax
  Upper Saddle River, (N.J.): Prentice Hall, 1996. [Online]. Available:
  \url{http://opac.inria.fr/record=b1091137}
\BIBentrySTDinterwordspacing

\bibitem{zuo2018three}
Z.~Zuo, L.~Cheng, X.~Wang, and K.~Sun, ``Three-dimensional path-following
  backstepping control for an underactuated stratospheric airship,'' \emph{IEEE
  Transactions on Aerospace and Electronic Systems}, vol.~55, no.~3, pp.
  1483--1497, 2018.

\bibitem{aguiar2003position}
A.~P. Aguiar and J.~P. Hespanha, ``Position tracking of underactuated
  vehicles,'' in \emph{American Control Conference, 2003. Proceedings of the
  2003}, vol.~3.\hskip 1em plus 0.5em minus 0.4em\relax IEEE, 2003, pp.
  1988--1993.

\bibitem{jain2018cooperative}
R.~P. Jain, A.~Alessandretti, A.~P. Aguiar, and J.~B. De~Sousa, ``Cooperative
  moving path following using event based control and communication,'' in
  \emph{2018 13th APCA International Conference on Automatic Control and Soft
  Computing (CONTROLO)}.\hskip 1em plus 0.5em minus 0.4em\relax IEEE, 2018, pp.
  189--194.

\bibitem{soetanto2003adaptive}
D.~Soetanto, L.~Lapierre, and A.~Pascoal, ``Adaptive, non-singular
  path-following control of dynamic wheeled robots,'' in \emph{Decision and
  Control, 2003. Proceedings. 42nd IEEE Conference on}, vol.~2.\hskip 1em plus
  0.5em minus 0.4em\relax IEEE, 2003, pp. 1765--1770.

\bibitem{lapierre2007nonlinear}
L.~Lapierre and D.~Soetanto, ``Nonlinear path-following control of an auv,''
  \emph{Ocean engineering}, vol.~34, no. 11-12, pp. 1734--1744, 2007.

\bibitem{cichella2011geometric}
V.~Cichella, I.~Kaminer, V.~Dobrokhodov, E.~Xargay, N.~Hovakimyan, and
  A.~Pascoal, ``Geometric 3d path-following control for a fixed-wing uav on so
  (3),'' in \emph{AIAA Guidance, Navigation, and Control Conference}, p. 6415.

\bibitem{cichella2013quadrotor}
\BIBentryALTinterwordspacing
V.~Cichella, R.~Choe, S.~B. Mehdi, E.~Xargay, N.~Hovakimyan, I.~Kaminer, and
  V.~Dobrokhodov, ``A 3d path-following approach for a multirotor uav on
  so(3),'' \emph{IFAC Proceedings Volumes}, vol.~46, no.~30, pp. 13 -- 18,
  2013, 2nd IFAC Workshop on Research, Education and Development of Unmanned
  Aerial Systems. [Online]. Available:
  \url{http://www.sciencedirect.com/science/article/pii/S1474667015402654}
\BIBentrySTDinterwordspacing

\bibitem{breivik2009guidance}
M.~Breivik and T.~I. Fossen, ``Guidance laws for autonomous underwater
  vehicles,'' \emph{Underwater vehicles}, vol.~4, pp. 51--76, 2009.

\bibitem{fossen2014uniform}
T.~I. Fossen and K.~Y. Pettersen, ``On uniform semiglobal exponential stability
  (usges) of proportional line-of-sight guidance laws,'' \emph{Automatica},
  vol.~50, no.~11, pp. 2912--2917, 2014.

\bibitem{caharija2016integral}
W.~Caharija, K.~Y. Pettersen, M.~Bibuli, P.~Calado, E.~Zereik, J.~Braga, J.~T.
  Gravdahl, A.~J. S{\o}rensen, M.~Milovanovi{\'c}, and G.~Bruzzone, ``Integral
  line-of-sight guidance and control of underactuated marine vehicles: Theory,
  simulations, and experiments,'' \emph{IEEE Transactions on Control Systems
  Technology}, vol.~24, no.~5, pp. 1623--1642, 2016.

\bibitem{sujit2014pathfollowing}
P.~B. Sujit, S.~Saripalli, and J.~B. Sousa, ``Unmanned aerial vehicle path
  following: A survey and analysis of algorithms for fixed-wing unmanned aerial
  vehicless,'' \emph{IEEE Control Systems}, vol.~34, no.~1, pp. 42--59, Feb
  2014.

\bibitem{tiago20173dMPF}
T.~Oliveira, A.~P. Aguiar, and P.~Encarnação, ``Three dimensional moving path
  following for fixed-wing unmanned aerial vehicles,'' in \emph{2017 IEEE
  International Conference on Robotics and Automation (ICRA)}, May 2017, pp.
  2710--2716.

\bibitem{wang2017mpf}
Y.~Wang and D.~Wang, ``Non-singular moving path following control for an
  unmanned aerial vehicle under wind disturbances,'' in \emph{2017 IEEE 56th
  Annual Conference on Decision and Control (CDC)}, Dec 2017, pp. 6442--6447.

\bibitem{rucco2015virtualtarget}
A.~Rucco, A.~P. Aguiar, and J.~Hauser, ``A virtual target approach for
  trajectory optimization of a general class of constrained vehicles,'' in
  \emph{2015 54th IEEE Conference on Decision and Control (CDC)}, Dec 2015, pp.
  5245--5250.

\bibitem{KAPITANYUK20176983}
Y.~A. Kapitanyuk, H.~G. de~Marina, A.~V. Proskurnikov, and M.~Cao, ``Guiding
  vector field algorithm for a moving path following problem,''
  \emph{IFAC-PapersOnLine}, vol.~50, no.~1, pp. 6983 -- 6988, 2017, 20th IFAC
  World Congress.

\bibitem{guan2019mpf}
Z.~{Guan}, Y.~{Ma}, and Z.~{Zheng}, ``Moving path following with prescribed
  performance and its application on automatic carrier landing,'' \emph{IEEE
  Transactions on Aerospace and Electronic Systems}, pp. 1--1, 2019.

\bibitem{zheng2020mpf}
\BIBentryALTinterwordspacing
Z.~Zheng, ``Moving path following control for a surface vessel with error
  constraint,'' \emph{Automatica}, vol. 118, p. 109040, 2020. [Online].
  Available:
  \url{http://www.sciencedirect.com/science/article/pii/S0005109820302387}
\BIBentrySTDinterwordspacing

\bibitem{jain2018auv}
R.~{Praveen Jain}, A.~{Pedro Aguiar}, and J.~B. d.~{Sousa}, ``Target tracking
  using an autonomous underwater vehicle: A moving path following approach,''
  in \emph{2018 IEEE/OES Autonomous Underwater Vehicle Workshop (AUV)}, Nov
  2018, pp. 1--6.

\bibitem{beard2012small}
R.~W. Beard and T.~W. McLain, \emph{Small unmanned aircraft: Theory and
  practice}.\hskip 1em plus 0.5em minus 0.4em\relax Princeton university press,
  2012.

\bibitem{bishop1975there}
R.~L. Bishop, ``There is more than one way to frame a curve,'' \emph{The
  American Mathematical Monthly}, vol.~82, no.~3, pp. 246--251, 1975.

\bibitem{hanson1995parallel}
A.~J. Hanson and H.~Ma, ``Parallel transport approach to curve framing,'' 1995.

\end{thebibliography}

\end{document}